\newif\iffigs\figstrue

\documentclass[paper, 12pt, letterpaper, epsf]{JHEP}
\input{epsf.tex}
\usepackage{graphics}
\def\Bbb{\bf}
\def\C{{\Bbb C}}
\def\R{{\Bbb R}}
\def\Z{{\Bbb Z}}

\def\bearray{\begin{eqnarray}}
\def\eearray{\end{eqnarray}}
\def\bearraynn{\begin{eqnarray*}}
\def\eearraynn{\end{eqnarray*}}
\def\bfig{\begin{figure}}
\def\efig{\end{figure}}

\def\opeq#1{\advance\lineskip#1 \advance\baselineskip#1
        \advance\lineskiplimit#1}

\newtheorem{Proposition}{Proposition}[section]

\newtheorem{Theorem}{Theorem}[section]
\newtheorem{Lemma}{Lemma}[section]
\newtheorem{Corrolary}{Corrolary}[section]

\newcommand{\be}{\begin{equation}}
\newcommand{\ee}{\end{equation}}
\newcommand{\bea}{\begin{eqnarray}}
\newcommand{\eea}{\end{eqnarray}}

\newcommand{\bp}{\begin{Proposition}}
\newcommand{\ep}{\end{Proposition}}
\newcommand{\bt}{\begin{Theorem}}
\newcommand{\et}{\end{Theorem}}
\newcommand{\bl}{\begin{Lemma}}
\newcommand{\el}{\end{Lemma}}
\newcommand{\bc}{\begin{Corrolary}}
\newcommand{\ec}{\end{Corrolary}}
\newcommand{\nn}{\nonumber}

\font\mybb=msbm10 at 12pt

\def\bb#1{\hbox{\mybb#1}}

\def\Z {\bb{Z}}

\def\id{\protect{{1 \kern-.28em {\rm l}}}}

\def\p#1{{\phi{}^{(#1)}}}
\def\hp#1{{{\phi'}{}^{(#1)}}}

\def \ov {\over }
\def\bea{\begin{eqnarray}}
\def\eea{\end{eqnarray}}

\def\k{{\bf k}}

\def\be{\begin{equation}}
\def\ee{\end{equation}}
\def\ba{\begin{eqnarray}}
\def\ea{\end{eqnarray}}

\def\1{{{(1)}}}\def\2{{{(2)}}}\def\3{{{(3)}}}

\font\mybb=msbm10 at 12pt

\def\bb#1{\hbox{\mybb#1}}

\def\Z {\bb{Z}}

\def\id{\protect{{1 \kern-.28em {\rm l}}}}

\def\p#1{{\phi{}^{(#1)}}}
\def\hp#1{{{\phi'}{}^{(#1)}}}

\usepackage{graphics}

\title{Gauge-fixing, semiclassical approximation and 
potentials for graded Chern-Simons theories}

\author{C.~I.~Lazaroiu
\\C.~N.~Yang Institute for Theoretical Physics\\
SUNY at Stony Brook, NY11794-3840,
U.S.A.\\calin@insti.physics.sunysb.edu\\
}
\author{R.~Roiban
\\Department of Physics, UCSB\\
Santa Barbara, CA 93106,
U.S.A.\\radu@vulcan.physics.ucsb.edu}

\abstract{We perform the Batalin-Vilkovisky analysis of gauge-fixing for 
graded Chern-Simons theories. Upon constructing an appropriate gauge-fixing 
fermion, we implement a Landau-type constraint, 
finding a simple form of the gauge-fixed action.  
This allows us to extract the associated Feynman rules
taking into account the role 
of ghosts and antighosts. Our gauge-fixing procedure 
allows for zero-modes, hence is not limited to the acyclic case.
We also discuss the semiclassical approximation and 
the effective potential for massless modes, thereby justifying 
some of our previous constructions in the  Batalin-Vilkovisky 
approach.}

\preprint{YITP-SB 01-63}

\begin{document}

\tableofcontents

\pagebreak

\vskip .6in

\section{Introduction}

The study of D-brane composites is central to a 
better understanding of Calabi-Yau compactifications of open strings
\cite{Douglas_Kontsevich, Aspinwall, Douglas_Aspinwall}.
In a series of papers \cite{com1, com3, sc, Diaconescu, bv, gauge, rs} 
it was proposed
that the topological counterpart of this problem (which can be 
formulated within the topological sigma model framework of 
\cite{Witten_nlsm, Witten_mirror}) can be studied with the 
tools of string field theory. This is a potentially fruitful approach,
since it allows us to apply standard field theory techniques to a 
seemingly unrelated problem. 

In particular, it was argued in \cite{sc} that the dynamics of
A-type topological branes in the absence of worldsheet instantons 
can be described by a graded version of Chern-Simons field theory, 
just as ungraded topological D-brane dynamics can be described in 
usual Chern-Simons language \cite{Witten_CS}. 
These models, whose equations of motion describe so-called 
`flat superconnections of total degree one', seem to allow for a standard 
description of the extended moduli space of topological D-branes. 
From a purely field theoretic perspective, they form interesting 
generalizations
of Chern-Simons field theory. It is therefore natural to ask how the 
results for the ungraded case extend to these more general systems.

In two recent papers \cite{gauge, rs}, we considered effective potentials
and the semiclassical approximation for such models. 
Since graded Chern-Simons theories contain higher rank forms, 
a full justification of the gauge-fixing procedure used in those papers 
requires a detailed analysis in the Batalin-Vilkovisky formalism. 
The purpose of the present paper is to carry this out in full generality, 
thereby justifying certain points of \cite{gauge,rs}
and complementing the classical BV analysis  already performed in \cite{bv}, 
upon using the geometric formalism of \cite{Witten_antibracket, 
Henneaux_geom, Khudaverdian, Schwarz_geom, Schwarz_semiclassical, 
Schwarz_symms, Schwarz_superanalogues, Kontsevich_Schwarz}. 
This will be achieved by considering a Landau-type gauge, 
which leads to a simple expression for the gauge-fixed
action reminiscent of the results of \cite{AS1}. 

Since we wish to work generally, the resulting BV analysis is somewhat 
technical. Indeed, a general graded Chern-Simons theory contains fields 
of arbitrarily large ghost number, even though the form rank is constrained 
to lie in the interval $0\dots 3$. The reason for this is that the ghost 
degree results by shifting the form rank of the field through a quantity which 
depends on position of the corresponding block inside a matrix of bundle-valued
forms. This has the effect of allowing an extended range of ghost numbers, 
which depends of the range of subbundle grades present in the system. 
To describe this situation, we will have to construct a gauge-fixing fermion 
which generally contains an arbitrary number of levels of extraghosts. 
This somewhat complicated structure can be described systematically upon 
using the general methods discussed in \cite{Gomis}, and leads to a 
result which is reminiscent of the `universal gauge
fermion' considered in a different context in \cite{Nash_Connor}.
Despite the complexity of this construction, the result of gauge-fixing 
is surprisingly simple, and in many ways similar to the 
situation familiar from usual Chern-Simons theory. Indeed, our 
gauge fermion will implement a graded analogue of the standard Landau-gauge, 
and lead to a description of the gauge-fixed 
action which is formally similar to that used (for the ungraded case) 
in \cite{AS1}. In fact, we shall be able to generalize even further, 
by including a convenient gauge-fixing procedure for zero modes 
(in particular, our construction is not limited to the acyclic case).
For the Landau-type fermion, the zero-mode components depend on certain 
parameters described by positive operators defined 
on the harmonic subspace
of various field, ghost and antighost 
configurations. This auxiliary data will be necessary in order to 
recover the description of the semiclassical approximation given 
in \cite{rs}. We also construct a family of weighted gauges, which reduces 
to the Landau gauge in an appropriate limit. 

An interesting by-product of gauge-fixing is that it induces certain prefactors
in the gauge-fixed path integral. These factors, which arise when 
integrating out the various auxiliary fields, are crucial for obtaining the 
correct path integral measure, and for recovering the 
results of \cite{rs} in the semiclassical approximation. 
In particular, we show that they are in complete 
agreement with the predictions of the method of resolvents 
\cite{Schwarz_resolvent, Schwarz_resolvent2, Adams_Sen}, 
which was used in \cite{rs}. 
In fact, we will be able to recover the results of \cite{rs} by a direct 
path integral computation. This agrees with the general observation 
\cite{Blau_Thompson,GK} that the method of resolvents is an indirect way of 
taking into account the role of ghosts and antighosts.
Finally, we give a BV treatment of the 
effective potential for massless modes, thus justifying certain statements of 
\cite{gauge}. 

The paper is organized as follows. In Section 2, we review the construction 
of graded Chern-Simons theories. We also discuss how one
can introduce a Hermitian structure on the space of field configurations, 
upon using auxiliary metric data which will be required by our gauge-fixing 
procedure. In Section 3, we review the tree-level BV action for our systems, 
which was constructed in \cite{sc} and discussed in detail in \cite{bv}. 
We also explain how one can promote the Hermitian structure on the 
space of classical 
fields to the space of extended field configurations 
(which includes configurations of ghosts and antifields).
Some of these issues have already been discussed 
in \cite{sc, bv} and \cite{gauge}, but we include a summary for 
reasons of completeness. In Section 4, we take up the problem of 
gauge-fixing in the BV formalism. Following standard procedure, we start 
in Subsection 4.1 
by adding trivial pairs of extraghosts and auxiliary fields. Subsection 
4.2. constructs the Landau gauge fixing fermion (a fermion of 
`delta-function type') upon using the general methods explained in 
\cite{Gomis}.  After describing the construction for both the harmonic and 
non-zero modes, we perform gauge-fixing by eliminating the antifields
and integrating out the auxiliary fields. This leads to our gauge-fixed 
action, and produces certain prefactors in the gauge-fixed path integral. 
The result of the entire process is summarized in Subsection 4.3. 
The gauge-fixed action has a particularly simple form, which is reminiscent of 
that considered in \cite{AS1} for the case of usual Chern-Simons theory. 
As explained in Subsection 4.4, this 
allows one to further simplify the gauge-fixed correlators, upon 
performing a certain change of variables for the antighosts. The result
is a graded version of the formalism used in \cite{AS1}, further extended 
to take into account the zero-modes. In Section 5, we use our results to study 
the semiclassical approximation and the effective potential. Upon 
considering a background field configuration, we expand the action in 
quadratic and cubic terms, and separate the result into the Gaussian 
contribution (which is given entirely by the massive modes) and 
the contributions coming from cubic terms. Up to a prefactor induced by the 
gauge-fixing of zero-modes, the former gives the semiclassical 
partition function, which is computed by two equivalent methods in Subsection 
5.1. Upon performing zeta-function regularization, the result is in complete 
agreement with that of \cite{rs}, which was obtained by the 
method of resolvents discussed in 
\cite{Schwarz_resolvent, Schwarz_resolvent2, Adams_Sen, 
Adams_Sen0}. The higher contributions 
can be used to define an effective potential for the zero-modes, which 
was already studied in \cite{gauge} with more elementary methods; this 
potential is induced due to the presence of interactions between harmonic 
and massive modes. 
In Section 6, we discuss the perturbative expansion of this potential, 
the relevant massive propagator and the tree-level approximation. For the 
latter, we explain the relation with the description of \cite{gauge}, and 
in particular justify the construction used in that paper from the 
BV perspective. Section 7 presents our conclusions. In the appendix, we
show that the Landau-type gauge of the present paper can be obtained as 
a certain limit of a family of weighted gauges, a result which generalizes 
well-known constructions of standard gauge theory. While conceptually 
important, this result is not 
needed for understanding the rest of the paper. The appendix also discusses a 
simple example, showing how the weighted gauge produced by our general 
construction can be re-discovered through more elementary means.

\section{Graded Chern-Simons theory}

We start with a brief review of graded Chern-Simons theories on closed 
3-manifolds. More details on their construction and basic properties 
can be found in \cite{gauge, bv}. The relation with graded topological D-branes
is explained in \cite{sc} (see also \cite{com1, com3}).

\subsection{The set-up}

Consider an oriented closed (i.e. compact and boundary-less) 3-manifold $L$,
and a (finite) collection of flat complex vector bundles $E_n$. 
We form the total bundle ${\bf E}=\oplus_{n}{E_n}$, endowed with the 
$\Z$-grading induced by $n$.  
We shall make the convention that a form on $L$ of rank lying outside 
the interval $0..3$ is defined to be zero. Note that we consider 
{\em complex} flat vector bundles $E_n$, which are not required to be unitary 
(i.e. there need not exist metrics on $E_n$ which are covariantly-constant 
with respect to the flat connections).

The graded Chern-Simons theory of \cite{sc, bv} describes sections  
of the bundle:
\be
{\cal V}=\Lambda^*(T^*L)\otimes End({\bf E})~~,~~
\ee
which we endow with the total grading 
${\cal V}=\oplus_{t}{{\cal V}^t}$, where:
\be
{\cal V}^t=\oplus_{\tiny \begin{array}{c}k, m, n\\k+n-m=t\end{array}}
{\Lambda^k(T^*L)\otimes Hom(E_m, E_n)}~~.
\ee
The space of sections ${\cal H}=\Gamma({\cal V})=\Omega^*(L, End({\bf E}))$  
is endowed with the grading ${\cal H}^k=\Gamma({\cal V}^k)$.
The degree of a section $u\in {\cal H}$ has the form:
\be
\label{deg}
|u|=rk u + \Delta(u)~~,
\ee
where $\Delta(u)=n-m$ if $u\in \Omega^*(L, Hom(E_m,E_n))$.

The action is most conveniently described in terms of the the so-called 
{\em total boundary product}:
\be
\label{bullet}
u\bullet v= (-1)^{\Delta(u)rk v}u\wedge v~~,
\ee
where the wedge product is understood to include composition 
of bundle morphisms. This associative product has the properties:
\be
\label{bullet_props}
|u\bullet v|=|u|+|v|~~, ~~1\bullet u=u\bullet 1=u~~,~~|u|=0~~,
\ee
where $1$ stands for the identity endomorphism of ${\bf E}$. 

The direct sum $A^{(0)}=\oplus_n{A_n}$ of the flat connections carried by $E_n$
induces a flat structure on $End({\bf E})$, and a differential $d^{(0)}$ 
(the de Rham differential twisted by this flat connection), which  
acts as a degree one derivation of the boundary product. 
More general backgrounds are obtained upon shifting by degree one elements 
$\phi$ of ${\cal H}$,  which leads to the shifted differential 
$d=d^{(0)}+[\phi,.]_\bullet$, 
where $[.,.]_\bullet$ stands for the graded 
commutator: 
\be
[u,v]_\bullet:=u\bullet v -(-1)^{|u||v|}v\bullet u~~.
\ee
Then $d$ is a degree one derivation of the associative algebra 
$({\cal H},\bullet)$:
\be
\label{d_props}
|du|=|u|+1~~,~~d(u\bullet v)=(du)\bullet v+(-1)^{|u|}u\bullet (dv)~~.
\ee
In the language of 
\cite{Bismut_Lott}, $d$ defines `graded superconnection \cite{Quillen} 
of total degree one'.
For what follows, we pick a reference background $\phi$ satisfying $d^2=0$. 

To write down the action, we also introduce the graded trace of elements 
$u$ in ${\cal H}$:
\be
\label{str}
str(u)=\sum_{n}{(-1)^n tr(u_{nn})}~~,~~{\rm for}~~u=\oplus_{m,n}{u_{mn}}~~,
\ee
with $u_{mn}\in \Omega^*(L, Hom(E_m, E_n))$. 
This allows us to define the nondegenerate bilinear form:
\be
\label{bf_A}
\langle u, v \rangle:=\int_{L}{str(u\bullet v)}~~, 
\ee
which has the properties:
\bea
\label{form_invariance}
\langle u, v\rangle =(-1)^{|u||v|}\langle v, u \rangle~~,~~
\langle du, v\rangle +(-1)^{|u|}\langle u, dv\rangle=0~~,~~
\langle u\bullet v, w\rangle =\langle u, v\bullet w\rangle~~ 
\eea
and obeys the selection rule $\langle u, v\rangle=0$ unless $|u|+|v|=3$.

\subsection{The action} 

The graded Chern-Simons theory is described by the action:
\be
\label{action}
S(\phi)=\int_{L}{str\left[\frac{1}{2}\phi\bullet d\phi+\frac{1}{3}
    \phi\bullet\phi\bullet\phi\right]} +cc=
\frac{1}{2}\langle \phi, d\phi\rangle +\frac{1}{3}\langle
\phi,\phi\bullet \phi\rangle +cc~~~~,  
\ee which is defined on the degree one subspace: 
\be
{\cal H}^1=\{\phi\in {\cal H}||\phi|=1\}=
\Gamma(\oplus_{k+n-m=1}{\Lambda^k(T^*L)\otimes Hom(E_m,E_n)})~~.
\ee 
The equations of motion have the form:
\be
\label{mc}
d\phi+\frac{1}{2}[\phi,\phi]_\bullet =0 \Leftrightarrow d\phi+
\phi\bullet \phi=0~~(\phi\in {\cal H}^1)~~,
\ee
and are equivalent with the requirement that the shifted 
superconnection $d_\phi=d+[\phi,.]_\bullet$ is flat 
(i.e. satisfies $(d_\phi)^2=0$). 
These equations are invariant under transformations of the form: 
\be
\label{Gauge}
\phi\rightarrow \phi^g=g\bullet \phi \bullet g^{-1}+g\bullet dg^{-1}~~,
\ee
where $g$ is an invertible element of the subalgebra $({\cal H}^0, \bullet)$.
Dividing the space of solutions to 
(\ref{mc}) through these symmetries gives a moduli space ${\cal M}$.

\subsection{Hermitian structure}

\subsubsection{Metrics} 

The gauge-fixing procedure discussed in Section 4 
will require choosing a Riemannian 
metric $g$ on $L$ and Hermitian metrics on the bundles
$E_n$. These a Hermitian metric $g_{\bf E}$ on ${\bf E}$, 
and thus a Hermitian metric on $End({\bf E})$:
\be
(\alpha,\beta)=tr(\alpha^\dagger\circ \beta)~~{\rm~for~}~~\alpha,\beta\in 
End({\bf E}_p)~~,p\in L~~,
\ee
where $\alpha^\dagger$ is the Hermitian conjugate of $\alpha$ with respect to 
$g_{\bf E}$. On the other hand, $g$ induces a Hermitian 
metric $(.,.)$ on $\Lambda^*(T^*L)$: 
\be
(*{\overline \omega})\wedge \eta=(\omega, \eta)vol_g~~,{\rm~for~}~~
\omega, \eta\in \Lambda^*(T^*_pL)~~,
\ee
where $vol_g$ is the volume form induced by $g$ on $L$ (with respect to the 
orientation on $L$), while $*$ is the {\em complex linear} Hodge operator, 
which satisfies:
\be
rk(*\omega)=3-rk\omega~~{\rm~and~}~~*^2=id~~.
\ee
Combining everything, 
we obtain a Hermitian metric $(.,.)_{\cal V}$ on the bundle 
${\cal V}=\Lambda^*(T^*L)\otimes End({\bf E})$:
\be
\label{Vmetric}
tr(*u^\dagger\wedge v)=(u,v)_{\cal V}
vol_g~~{\rm~for~}u,v\in {\cal V}_p~~,~~p\in L~~.
\ee
Integration over $L$ gives a Hermitian scalar product on the 
space ${\cal H}=\Gamma({\cal V})$:
\be
\label{hmetric}
h(u,v)=\int_{L}{(u,v)vol_g}=
\int_{L}{tr(*u^\dagger\wedge v)}~~,~~{\rm for}~~u,v\in {\cal H}~~.
\ee

\subsubsection{The conjugation operator} 

Since the bilinear form (\ref{bf_A}) is non-degenerate, there exists a unique 
antilinear operator $c$ on ${\cal H}$ with the property:
\be
\label{h}
h(u,v)=\langle cu,v\rangle=\int_{L}{str[(cu)\bullet v]}~~.
\ee
This has the following form on decomposable elements:
\be
\label{c_A}
c(\omega\otimes f)=(-1)^{n+\Delta(f)(1+rk \omega)}
(*{\overline \omega})\otimes f^\dagger~~,~~{\rm~for~}~~\omega\in \Omega^*(L)~~
{\rm~and~}~~f\in Hom(E_m,E_n)~~
\ee
and satisfies:
\be
\label{c_props}
|cu|=3-|u|~~, ~~c^2=id~~,~~h(u,v)=h(cv,cu)
\ee
(in particular, $c$ is anti-unitary). 
In the ungraded case (${\bf E}=E_0$), 
$c$ reduces to the {\em antilinear} Hodge operator 
${\overline *}$, coupled to the bundle $End(E)$.

\subsubsection{The adjoint of $d$ and the Laplacian}

The scalar product $h$ satisfies the selection rule:
\be
\label{h_sel}
h(u,v)=0 ~~,~~{\rm~unless~}~~|u|=|v|~~.
\ee
Considering the Hermitian conjugate $d^\dagger$ of $d$ (with respect to $h$), 
one has:
\be
\label{dder}
d^\dagger u= (-1)^{|u|}cdcu~~,~~|d^\dagger u|=|u|-1~~,~~(d^\dagger)^2=0~~,
~~\langle d^\dagger u, v\rangle =(-1)^{|u|}\langle u, d^\dagger v\rangle~~.
\ee
and:
\bea
\label{cdd}
cd^\dagger u= (-1)^{|u|}dcu~~&,&~~d^\dagger cu=(-1)^{|u|+1}cdu~~\nn\\
d^\dagger dc=cdd^\dagger~~&,&~~dd^\dagger c=cd^\dagger d~~.
\eea
One also constructs the `deformed Laplacian' 
$\Delta=dd^\dagger+d^\dagger d$, which will be useful below.

We recall the Hodge decompositions:
\be
{\cal H}=K\oplus im d \oplus im d^\dagger~~,~~ker d=imd \oplus K~~,
~~ker d^\dagger= im d^\dagger \oplus K~~,
\ee
where $K=ker \Delta=ker d\cap ker d^\dagger$.
It will be useful to 
consider the orthogonal projectors $\pi_{d}=d\frac{1}{\Delta}d^\dagger, 
\pi_{d^\dagger}=d^\dagger \frac{1}{\Delta}d$ and 
$P=1-\pi_d-\pi_{d^\dagger}$ of ${\cal H}$ on 
the subspaces $im d$, $im d^\dagger$ and $K$, 
as well as the propagator $U:=\frac{1}{d}\pi_d$. 
The first of equations (\ref{cdd}) implies the relation:
\be
\pi_d c=c\pi_{d^\dagger}~~.
\ee

\subsection{Spacetime ghost grading}

For what follows, it will be convenient to use the modified grading 
$s(u)=1-|u|$, which in the 
BV formalism corresponds to the ghost number 
of the string field theory \cite{bv}. 
We shall use the notation ${\cal H}(\sigma)={\cal H}^{1-\sigma}$ 
for the homogeneous subspaces of ${\cal H}$ with respect to this grading, 
and the notation $H_\sigma({\cal H})=H^{1-\sigma}({\cal H})$ for the 
associated components of $H^*({\cal H})$ (note that ghost grading 
is decreased by $d$, which justifies the homological notation).
With this convention, the string field lies in the subspace ${\cal H}(0)$. 
One has:
\be
\label{s_props}
s(du)=s(u)-1~~,~~s(cu)=-1-s(u)~~,~~s(cdu)=-s(u)~~.
\ee
In particular, $cd$ induces an antilinear operator on the 
physical subspace ${\cal H}(0)$. 

\subsection{Dependence on the underlying superbundle}

It is instructive to consider to what extend the various data are specified 
by the superbundle underlying our graded bundle. For this, we 
define subbundles:
\be
E_{even}=\oplus_{n=even}{E_n}~~{\rm and}~~E_{odd}:=\oplus_{n=odd}{E_n}~~,
\ee
so that ${\bf E}=E_{even}\oplus E_{odd}$. Viewed in this manner, ${\bf E}$ 
becomes a superbundle(=$\Z_2$-graded bundle), if one forgets the 
finer $\Z$-grading given by the decompositions of $E_{even}$ and $E_{odd}$. 
Accordingly, the bundle ${\cal V}$ only remembers the $\Z_2$-grading
${\cal V}={\cal V}^{even}\oplus {\cal V}^{odd}$, where: 
\be
{\cal V}^{even}:=\oplus_{k=even}{{\cal V}^k}~~{\rm and}~~
{\cal V}^{odd}:=\oplus_{k=odd}{{\cal V}^k}~~.
\ee
These subbundles only depend on the superbundle structure of ${\bf E}$:
\bea
{\cal V}^{even}=\oplus_{j=even}{\left[
\Omega^{j}(L, Hom(E_{even},E_{even}))\oplus
\Omega^{j}(L, Hom(E_{odd},E_{odd}))\right]}\oplus\nn\\
\oplus_{j=odd}{\left[\Omega^{j}(L, Hom(E_{even},E_{odd}))\oplus
\Omega^{j}(L, Hom(E_{odd},E_{even}))\right]}~~,
\eea
and a similar relation for ${\cal V}^{odd}$. Moreover, the $\Z$-grading 
on ${\cal H}$ induces a $\Z_2$-grading which 
is only sensitive to the superbundle structure of ${\bf E}$:
\be
{\cal H}^{even}:=\oplus_{k=even}{{\cal H}^k}=\Gamma({\cal V}^{even})~~,~~
{\cal H}^{odd}:=\oplus_{k=odd}{{\cal H}^k}=\Gamma({\cal V}^{odd})~~.
\ee
With respect to this $\Z_2$-grading, the boundary algebra 
$({\cal H}, d,\bullet)$ becomes a differential superalgebra. 
The essential point, however, is that this description {\em does not 
suffice in order to identify the physical fields}, and in particular does 
not uniquely specify the physical theory. Indeed, the physical subspace 
${\cal H}^1=\{\phi\in {\cal H}||\phi|=1\}$ {\em knows} about the $\Z$-grading 
of ${\bf E}$. The same is true about the gauge-group (\ref{Gauge}), whose 
structure depends markedly of this grading. Therefore, the physical content 
of the theory is entirely different for 
various choices of $\Z$-grading compatible with a given $\Z_2$-grading. 
This becomes especially clear when one considers the moduli space, 
whose structure is very sensitive to choice of  a $\Z$-valued grading. 
Accordingly, the nature of the fields one can condense depends on the precise 
choice of ghost grading, since it is this data which specifies which fields 
are physical. We also stress that our theories are quite different from the
super-Chern-Simons theories of \cite{supergroupCS, Vafa_cs}. 
In fact, the latter only contain 
physical {\em one-form} fields, while our theories will 
generally contain physical 
fields of rank different from one when both $E^{even}$ and $E^{odd}$ 
are non-vanishing. It is precisely for this reason that condensation 
of fields which are not one-forms can be achieved in our framework, 
thereby allowing for an interpretation in terms of topological 
D-branes \cite{sc}.

\section{The classical BV system}

\subsection{The tree-level master action}

Let us begin by recalling some 
results\footnote{The construction presented here (following \cite{bv})  
is carried out within the framework of \cite{DeWitt}. 
An alternate (and possibly better) point of view is the 
Berezin theory of superschemes \cite{Berezin}. We prefer to use the
formulation of \cite{DeWitt, Rogers} due to its being better 
established in the physics literature.} 
of \cite{bv}, which will necessary below. 
Since the gauge algebra of (\ref{action}) is generally reducible, 
a proper formulation of our theories requires the BV formalism 
already at the classical level. The corresponding classical 
master action was 
constructed in \cite{sc,bv}. The conclusion is as follows. The tree-level BV 
action associated with (\ref{action}) is given by:
\be
\label{extended_action}
S_e({\hat \phi})= 2 Re\int_L{ str_e\left[\frac{1}{2}{\hat \phi} *d{\hat
      \phi}+ \frac{1}{3}{\hat \phi}*{\hat \phi}*{\hat \phi}
      \right]}=Re\left[\langle {\hat \phi}, d{\hat \phi}
\rangle_e +\frac{2}{3}\langle {\hat \phi},{\hat \phi}* {\hat \phi}
\rangle_e\right]~~,  
\ee 
where the {\em extended field} ${\hat \phi}$ is an element of the 
subspace $M:={\cal H}_e^1=\{ {\hat \phi}\in {\cal H}_e |
deg {\hat \phi}={\hat 1}\}$ of the so-called {\em extended boundary space} 
${\cal H}_e$. The latter is constructed as the tensor product:
\be
{\cal H}_e={\cal H}\otimes G~~,
\ee
where $G$ is a (complex) auxiliary Grassmann algebra. As in \cite{bv}, we use $g$ 
to denote the $\Z_2$-degree on $G$ and $deg$ to denote the induced $\Z_2$-valued
degree on ${\cal H}_e$:
\be
deg(u\otimes \alpha)=|u|~(mod~2)~+g(\alpha)~~,
\ee
where $u\in {\cal H}$ and $\alpha\in G$. When endowed with this grading,
the extended boundary space is a differential superalgebra with respect 
to the extended boundary product $*$ defined through:
\be
\label{star}
(u\otimes \alpha)*(v\otimes \beta)=(-1)^{|v|g(\alpha)}
(u\bullet v )\otimes (\alpha\beta)~~
\ee 
and the differential $d_e=d\otimes id_G$. We also extend the gradings $|.|$ 
and $g$ to partial gradings on ${\cal H}_e$ by:
\be
|u\otimes \alpha|:=|u|~~{\rm and}~~g(u\otimes \alpha):=g(\alpha)~~,
\ee
so that $deg {\hat u}=|{\hat u}|~(mod ~2~)+g({\hat u})$ for all  ${\hat u}$ 
in ${\cal H}_e$. 

The $G$-valued, complex-bilinear
form $\langle ., .\rangle_e$ 
appearing in (\ref{extended_action}) is defined through
(for decomposable elements 
${\hat u}=u\otimes \alpha$ and ${\hat v}=v\otimes \beta$):
\be
\label{extended_form}
\langle u\otimes \alpha, v\otimes \beta\rangle_e=
(-1)^{|v|g(\alpha)}\langle u,v\rangle \alpha\beta=
\int_L{str_e({\hat u}*{\hat v})}~~,
\ee
with the extended supertrace given by: 
\be
str_e(\omega\otimes f\otimes \alpha)=str(f)\omega\otimes \alpha~~,
\ee
for $\omega\in \Omega^*(L)$, $f\in End({\bf E})$ and $\alpha\in G$. 

The ghost grading of the BV formalism is given by:
\be
s({\hat u}):=1-|{\hat u}|~~,
\ee
and leads to a decomposition ${\cal H}_e=\oplus_{\sigma}{{\cal H}_e(\sigma)}$, 
where ${\cal H}_e(\sigma):={\cal H}^{1-\sigma}\otimes G$ 
is the subspace of elements of 
ghost number equal to $\sigma$. Accordingly, the odd subspace 
$M={\cal H}^1_e$  decomposes as:
\be
M=\oplus_{\sigma}{M(\sigma)}~~,~~{\rm with}~~M(\sigma)=
{\cal H}_e^1\cap {\cal H}_e(\sigma)=\{{\hat \phi}\in {\cal H}_e|s({\hat \phi})=
\sigma~{\rm~and~}~deg {\hat \phi}={\hat 1}\}~~.
\ee
The extended field has the decomposition:
\be
\label{phi_decomp}
{\hat \phi}=\oplus_{\sigma}{{\hat \phi}_\sigma}=
\oplus_{\sigma\geq 0}{\phi^*_\sigma}\oplus
\oplus_{\sigma\geq 0}{\phi_\sigma}~~,~~{\rm with}~~
s({\hat \phi}_\sigma)=\sigma\Leftrightarrow {\hat \phi}_\sigma\in M(\sigma)~~,
\ee
where we introduced the notations 
${\hat \phi}_\sigma =\phi_\sigma$ for $\sigma \geq 0$ and
${\hat \phi}_{\sigma}=\phi_{-1-\sigma}^*$ for $\sigma<0$. 
The component ${\hat \phi}_0=\phi_0\in M(0)$ is the classical field, 
related to the unextended field $\phi$ of  Section 2 through:
\be
{\tilde ev}_G(\phi_0)=\phi~~,
\ee 
where ${\tilde ev}_G:=id\otimes ev_G:{\cal H}_e\rightarrow {\cal H}$
is the extension of the obvious evaluation map
$ev_G:G\rightarrow {\bf C}$. One has ${\tilde ev}_G(M(0))={\cal H}^1$.
The components 
$\phi_\sigma$ ($\sigma>0$) play the role of ghosts, while $\phi^*_\sigma$ 
$(\sigma \geq 0)$ are the antifields. It will aso be convenient to 
describe the collections of fields and antifields 
by the elements ${\hat \phi}_+:=\oplus_{\sigma\geq 0}{{\hat \phi}_\sigma}$ 
and ${\hat \phi}_{-}:=\oplus_{\sigma<0}{{\hat \phi}_\sigma}$, such 
that ${\hat \phi}:={\hat \phi}_{-}\oplus {\hat \phi}_{+}$. 

The $G$-valued extended action (\ref{extended_action}) 
relates to the classical action (\ref{action}) as follows (figure 1):
\be
\label{restriction} 
ev_G(S_e({\hat \phi}))=S({\tilde ev}_G({\hat \phi}))~{\rm~for~} {\hat \phi}\in
M(0):=\{{\hat \phi}\in {\cal H}_e^1||{\hat \phi}|=1\}~~.  
\ee
The restriction of $*$ 
to the subspace $M(0)$ coincides with the unextended product 
$\bullet$ up to application of ${\tilde ev}_G$:
\be
{\tilde ev}_G({\hat u}*{\hat v})={\tilde ev}_G(u)\bullet {\tilde ev}_G(v)~~,
~~{\rm~for~}~~{\hat u}~{\rm and}~{\hat v}\in M(0)~~.
\ee
We refer the reader to \cite{bv} for more details on the extended boundary 
data and the associated BV system.

\begin{figure}[hbtp]
\begin{center}
\scalebox{0.5}{\input{evaluation.pstex_t}}
\end{center}
\caption{Relation between the extended and unextended actions.}
\end{figure}

\subsection{Extended Hermitian product and conjugation}

As in Subsection 2.2., let us pick metrics on $L$ and $E_n$. We wish to 
extend the Hermitian data of that section to the space ${\cal H}_e$. 
For this, we shall
assume that the Grassmann algebra $G$ is endowed with a
complex conjugation, i.e. a complex-antilinear, involutive operator
$\overline{\cdot}:G\rightarrow G$ which satisfies:
\be
\overline{\alpha\beta}=
\overline{\beta}\overline{\alpha}~~,~~g(\overline{\alpha})=g(\alpha)~~.
\ee 
This allows us to define the {\em extended conjugation 
operator} $c_e:{\cal H}_e\rightarrow {\cal
H}_e$ via:
\be
c_e(u\otimes \alpha)=(-1)^{g(\alpha)(|u|+1)}c(u)\otimes \overline{\alpha}~~.
\ee
It is clear that $c_e$ is complex-antilinear and satisfies:
\be
|c_e({\hat u})|=3-|{\hat u}|~~,~~ g(c_e({\hat u}))=g({\hat u})~~,
~~deg c_e({\hat u})={\hat 1}+deg {\hat u}~~.
\ee
A simple computation shows that:
\be
c_e^2({\hat u})=(-1)^{g({\hat u})}{\hat u}~~.
\ee

We also define an {\em extended Hermitian product} through:
\be
h_e(u\otimes \alpha, v\otimes \beta)=h(u,v){\overline \alpha}\beta~~.
\ee
This Grassmann-valued pairing has the property:
\be
\label{herm_e}
\overline{h_e({\hat u}, {\hat v})}=h_e({\hat v}, {\hat u})~~.
\ee
Moreover, it is easy to check that:
\be
h_e({\hat u}, {\hat v})=\langle c_e{\hat u}, {\hat v}\rangle_e~~,
\ee
which parallels the defining relation of the unextended conjugation $c$.
The hermicity property (\ref{herm_e}) also reads
\be
\langle c_e{\hat u},{\hat v}\rangle_e=
\overline{\langle c_e{\hat v},{\hat u}\rangle_e}~~,  
\ee
a relation which will be useful below.

\subsection{The extended Laplacian}

If we define $d_e^\dagger=d^\dagger\otimes id_G$, then a simple computation
gives: 
\be
d_e^\dagger{\hat u}=(-1)^{|{\hat u}|}c_ed_ec_e{\hat u}~~,~~h_e({\hat u}, d_e{\hat v})=
h_e(d_e^\dagger {\hat u}, {\hat v})~~.
\ee
The {\em extended Laplacian} 
$\Delta_e=d_e^\dagger d_e+d_ed_e^\dagger=\Delta\otimes id_G$ 
is Hermitian in the following sense:
\be
h_e(\Delta_e{\hat u}, {\hat v})=h_e({\hat u}, \Delta_e{\hat v})~~.
\ee
It is easy to check that the last equation of (\ref{dder}) implies:
\be
\label{dagger_der}
\langle d_e^\dagger {\hat u}, {\hat v}\rangle_e=(-1)^{deg {\hat u}}
\langle {\hat u}, d_e^\dagger {\hat v}\rangle_e~~.
\ee
One can also check the relation:
\be
\label{cddc}
\langle {\hat u}, d_ec_e{\hat v}\rangle_e=(-1)^{(deg {\hat u}+1)deg {\hat v}}
\overline{\langle c_ed_e{\hat u}, {\hat v}\rangle_e}~~,
\ee
which will be useful in Section 4. 

As in Subsection 2.2.2, we consider the orthogonal
projectors $\pi_{d_e}=d_e\frac{1}{\Delta_e}d_e^\dagger=\pi_d\otimes id_G$,
$\pi_{d_e^\dagger}=d_e^\dagger\frac{1}{\Delta_e}d_e=\pi_{d^\dagger}
\otimes id_G$ 
and $P_e=1-\pi_{d_e}-\pi_{d_e^\dagger}=P\otimes id_G$
of ${\cal H}_e$ 
on the subspaces $im d_e=(im d) \otimes G$, 
$im d_e^\dagger=(im d^\dagger) \otimes G$ and 
$K_e:=kerd_e\cap kerd_e^\dagger= K\otimes G$. We
have:
\bea
\label{cdd_e}
c_ed_e^\dagger {\hat u}=(-1)^{deg {\hat u}}d_ec_e{\hat u}~~&,&~~
d_e^\dagger c_e{\hat u}=(-1)^{1+deg u}c_ed_e{\hat u}~~\nn\\
d_e^\dagger d_e c_e =c_ed_ed_e^\dagger ~~&,&~~
d_ed_e^\dagger c_e =c_ed_e^\dagger d_e~~
\eea 
which generalizes equations (\ref{cdd}).
The first relations in (\ref{cdd_e}) imply: 
\be
c_e\pi_{d_e^\dagger}=\pi_{d_e} c_e~~.
\ee
We further note that the invertible operator $c_e$
maps $im d_e^\dagger$ into $im d_e$ and viceversa. As in \cite{superpot},
$d_e^\dagger$ gives a bijection between $im d_e$ and $im
d_e^\dagger$, which means that $c_ed_e$ 
is invertible as an operator form $im d_e^\dagger$ to itself.

We end with an observation which will be useful in the
next section. If $f$ is a $G$-valued function, then we define its
`real part' by $Ref:=\frac{1}{2}(f+\overline{f})$.
$Ref$ is $G$-valued and has the property
$\overline{Ref}=Ref=Re\overline{f}$.
If ${\hat v}$ is a fixed element of ${\cal H}_e$, then the linear functional:
\be
\eta_{\hat v}({\hat u}):=Re\langle {\hat v}, {\hat u}\rangle_e
\ee
is differentiable at the origin, with differential:
\be
\label{diff}
d_0\eta_{\hat v}({\hat w})=Re \langle {\hat v}, {\hat w}\rangle_e~~. 
\ee
It is clear that $d_e\eta_v$ vanishes if and only if $\langle
{\hat w}, {\hat v}\rangle_e=0$ for all ${\hat w}\in {\cal H}_e$; this
follows upon substituting ${\hat w}$ by $i{\hat w}$
and considering both equations. Since the unextended bilinear form is 
non-degenerate, this implies vanishing of ${\hat v}$.

If $F$ is a $G$-valued functional of ${\hat u}\in {\cal H}_e$, then we define 
the (generally non-linear) operator 
$\frac{\delta F}{\delta {\hat u}}:{\cal H}_e\rightarrow {\cal H}_e$ by: 
\be
\label{delta}
\delta F(\delta {\hat w})=\langle \frac{\delta F}{\delta {\hat u}} ({\hat u}), 
\delta {\hat w}\rangle_e=
Re \langle \frac{\delta F}{\delta {\hat u}} ({\hat u}), 
\delta {\hat w}\rangle_e~~,
\ee
where we used the fact that $\delta F(\delta {\hat w})=
Re \delta F(\delta {\hat w})$. 
With this definition, one has 
$\frac{\delta \eta_v}{\delta {\hat u}}={\hat v}$, which is constant on 
${\cal H}_e$.

\section{Gauge fixing in the BV formalism}

We are now ready to perform the BV analysis of gauge-fixing. This will
serve as justification for the effective
potential discussed in \cite{gauge} and completes the 
classical BV treatment of \cite{bv}. It will also 
allow us to recover the semiclassical approximation of \cite{rs}
through a purely path integral approach.
As a by-product, we shall find an invariant expression 
for the propagators of physical fields and ghost/antighosts. 
While our description entails a certain level of abstraction, we shall be able 
to give an entirely general discussion of graded Chern-Simons systems, 
which is valid for an arbitrary number of flat bundles $E_n$. 
Since the next two subsections are somewhat technical, the 
casual reader can read this introduction and jump directly to 
Subsection 4.3, which summarizes the results. 
The contents of that subsection suffice for understanding 
the rest of the paper. 

Before proceeding with the technical details, let us explain the main 
points of our procedure. We wish to fix a background flat 
superconnection (i.e. a solution of the equations of motion) and 
build a local description of our theory around that background. 
In general, the background superconnection will not be isolated. 
Let $d$ denotes the differential of Section 2 in such a background
(note that $d$ does not arise from the original flat connections, but it 
is `twisted' with the background superconnection, in spite of our simplified 
notations). Then the linearized approximation to the moduli problem 
(\ref{mc}, \ref{Gauge}) shows that infinitesimal deformations of the 
background are described (in first approximation) by the 
cohomology space $H^1_d({\cal H})$; this description will be corrected at 
higher orders due the fact that some linearized 
deformations will be obstructed, 
so the space $H^1_d({\cal H})$ describes {\em virtual} 
moduli (obstructions will be described by the effective potential, 
as in \cite{gauge}). If the first cohomology of $d$ 
does not vanish, then one has to perform the path integral in the presence 
of zero modes. The standard description in this situation is to 
separate massive fluctuations around the background, and to decompose the 
path integral into an ordinary 
integral over zero modes (i.e. over the moduli
space) and a path integral over the massive modes. Since we are interested 
in a local description, we shall proceed in a slightly different manner.
Namely, we shall treat the zero modes in the linearized approximation (i.e. as
{\em virtual} moduli), while treating the nonzero-modes exactly. The effect 
of this will be to produce a potential for the zero modes (obtained by 
integrating out the massive fluctuations), which in turn allows one to 
describe obstructions to infinitesimal deformations, and thus give an 
equivalent local formulation of the moduli problem \cite{superpot, gauge}. 

Since we do not wish to integrate over (virtual) zero modes, we shall pick 
a gauge-fixing procedure which freezes these to some particular values.
More precisely, we shall pick auxiliary metrics and 
describe elements in $H^1_d({\cal H})$ 
by the harmonic 
component $\phi^K$ of the physical field $\phi$, upon using the 
Hodge-theoretic decomposition discussed in Section 2. Our gauge-fixing fermion
will freeze the virtual zero mode $\phi^K$ to some value $\phi^H$, which 
plays the role of a parameter in the gauge fermion. We shall 
allow nonzero values of $\phi^H$, since we wish to build an effective potential
for the virtual zero modes, hence we must allow them to take 
different values. 
In fact, $\phi^H$ should be viewed as an (infinitesimal) {\em potential} 
deformation of the background superconnection -- which will be obstructed or 
not, depending on whether it lies along a valley of the effective potential
or fails to be contained in its critical set. 

This basic picture must of course be modified by taking into account the 
ghosts and antighosts required for fixing the gauge symmetries of the 
original action. Since these are partners of the various components of the 
physical field, we shall treat them in a similar manner. Namely, we 
decompose all ghosts and antighosts into harmonic (massless) 
and nonharmonic (massive) components, and treat them separately in the 
gauge-fixing fermion. As for the physical field, the massless 
ghost/antighost modes will be fixed to some arbitrary values 
(we shall later take these to be 
zero, so that only the physical harmonic mode $\phi^H$ survives as 
a non-vanishing parameter). For the massive modes, we pick a Landau 
gauge condition, which reproduces the Lorentz gauge used in 
\cite{gauge}. This will be implemented by a `universal' gauge-fixing fermion
of delta-function type, built according to the general rules explained in 
\cite{Gomis}. Its construction is performed in standard manner, after 
adding an appropriate collection of trivial pairs to the 
original set of fields and antifields.

\subsection{Trivial pairs}

To perform gauge-fixing of the BV action (\ref{extended_action}), we 
must introduce an appropriate number of trivial pairs  
and pick a convenient gauge-fixing fermion \cite{Gomis}. Since we wish 
to do this in general, it will prove convenient to use the following 
notation.

Let $\phi[0]={\hat \phi}^+=\oplus_{\sigma \geq 0}{\phi_\sigma}$ 
and $\phi^*[0]={\hat \phi}_-=\oplus_{\sigma\geq 0}{\phi^*_\sigma}$
be the original collections of BV fields and antifields.
We introduce new fields $\phi[k]=\oplus_{\sigma\geq 2k}
{\phi_\sigma[k]}\in {\cal H}_e^1$ (for $k\geq 1)$ and ${\overline
\phi}[k]=\oplus_{\sigma\geq 2k+1}{ {\overline \phi}_\sigma[k]}
\in {\cal H}_e^1$ (for $k\geq 0$), such that:
\bea
s(\phi_\sigma[k])=s({\overline\phi}_\sigma[k])=\sigma~~&,
&~~ g(\phi_\sigma[k])=g({\overline \phi}_\sigma[k])=\sigma~(mod~2)\nn\\
gh(\phi_\sigma[k])=\sigma-2k~~&,&~~gh({\overline\phi}_\sigma[k])=2k-\sigma~~,
\eea
for all $k\geq 0$, where $gh$ stands for the ghost number. 
We have $gh(\phi_\sigma[k])+gh({\overline \phi}_\sigma[k])=0$
for all $k,\sigma$. Note that the ghost number $gh$ coincides with 
$s$ only for the original BV fields $\phi_\sigma[0]=\phi_\sigma$.  
The interpretation of the various fields is as
follows.  Remember that $\phi_0[0]=\phi_0$ is the physical field, while 
$\phi_\sigma[0]$ with $\sigma \geq 1$ are the ghosts. 
Then ${\overline\phi}_\sigma[0]$ (with $\sigma\geq 1$) are antighosts for
these ghosts. 
The fields $\phi_\sigma[1]$ ($\sigma\geq 2$) are
first level extraghosts, i.e. ghosts for the antighosts 
${\overline\phi}_\sigma[0]$ with $\sigma \geq 2$ (note that there is no 
extraghost  associated with the first antighost ${\overline \phi}_1[0]$).
Moreover, ${\overline\phi}_\sigma[1]$ 
($\sigma\geq 3$) are
antighosts for those extraghosts $\phi_\sigma[1]$ which have $\sigma\geq 3$
(again the first extraghost $\phi_2[1]$ is left unpaired). At the second 
level, we introduce new extraghosts $\phi_\sigma[2]$ and their antighosts 
${\overline \phi}_\sigma[2]$ and so on for higher levels. 
In general, ${\overline\phi}_\sigma[k]$ are antighosts
for the $k^{th}$ level extraghosts $\phi_\sigma[k]$ which have 
$\sigma\geq 2k+1$.  These fields can be
arranged in a triangular field diagram \cite{Gomis}, as shown in
figure 2. This diagram will contain a finite number of nodes for a system
based on a finite number of flat bundles $E_n$. 

\begin{figure}[hbtp]
\begin{center}
\mbox{\epsfxsize=10cm \epsffile{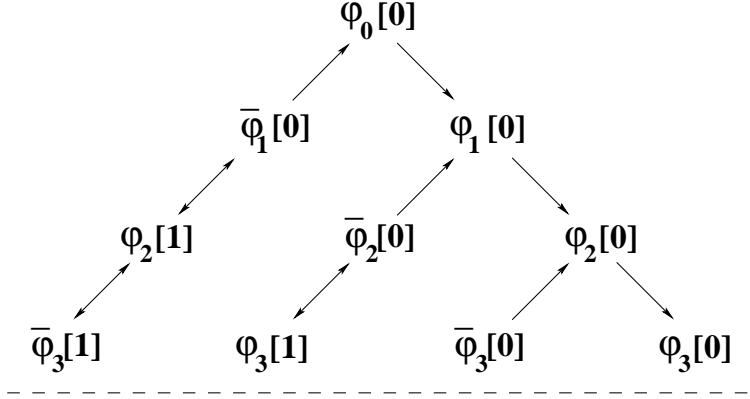}}
\end{center}
\caption{The triangular field diagram.}
\end{figure}

We also have the associated antifields $\phi^*[k]=\oplus_{\sigma\geq
2k}{\phi^*_\sigma[k]}\in {\cal H}_e^1$  
and ${\overline \phi}^*[k]=\oplus_{\sigma\geq
2k+1}{{\overline \phi}^*_\sigma[k]}\in {\cal H}_e^1$ (for $k\geq 0$), with:
\bea
s(\phi^*_\sigma[k])=s({\overline
\phi}_\sigma^*[k])=-\sigma-1~~&,&~~g(\phi^*_\sigma[k])=g({\overline
\phi}_\sigma^*[k])=\sigma+1~(mod~2)~~\nn\\ 
gh(\phi^*_\sigma[k])=2k-\sigma-1~~&,&~~
gh({\overline\phi}^*_\sigma[k])=\sigma-2k-1~~,
\eea
for all $k\geq 0$. Note that $\phi_\sigma^*[0]=\phi_\sigma^*$ are the
original antifields. Finally, we introduce auxiliary
fields $\pi[k]=\oplus_{\sigma\geq 2k}{\pi_\sigma[k]}\in {\cal H}^0$ 
(for $k\geq 1$) and 
${\overline \pi}[k]=\oplus_{\sigma\geq 2k+1}{{\overline \pi}_\sigma[k]}
\in {\cal H}^0$ 
(for $k\geq 0$) and their antifields
$\pi^*[k]=\oplus_{\sigma\geq 2k}{\pi^*_\sigma[k]}\in {\cal H}^0$ 
(for $k\geq 1$) and ${\overline \pi}[k]=\oplus_{\sigma\geq 2k+1}{
{\overline \pi}^*_\sigma[k]}\in {\cal H}^0$ 
(for $k\geq 0$), such that:
\bea
s(\pi_\sigma[k])=s({\overline \pi}_\sigma[k])=\sigma~~&,&
~~g(\pi_\sigma[k])=g({\overline \pi}_\sigma[k])=\sigma +1~(mod~2)~~\nn\\
gh(\pi_\sigma[k])=1+\sigma-2k~~&,&~~
gh({\overline \pi}_\sigma[k])=1-\sigma+2k~~,
\eea
and:
\bea
s(\pi^*_\sigma[k])=s({\overline \pi}^*_\sigma[k])=-1-\sigma~~&,&
~~g(\pi^*_\sigma[k])=g({\overline \pi}^*_\sigma[k])=\sigma~(mod~2)~~\nn\\
gh(\pi^*_\sigma[k])=2k-\sigma-2~~&,&~~ gh({\overline \pi}^*_\sigma[k])=
\sigma-2k-2~~.
\eea
The fields $(\pi_\sigma[k],\phi_\sigma[k])$ (for $k\geq 1$) and $({\overline
\phi}_\sigma[k],{\overline \pi}_\sigma[k])$ (for $k\geq 0$) 
form so-called {\em trivial pairs}
\cite{Gomis}. 

It will convenient for what follows to decompose all fields into 
harmonic and nonharmonic components, according to the Hodge decomposition:
\be
{\cal H}_e=N_e\oplus K_e~~{\rm~where}~~N_e=im d_e\oplus im d_e^\dagger~~.
\ee
Thus, we write: 
\bea
\phi_\sigma[k]=\phi_\sigma^N[k]\oplus \phi_\sigma^K[k]~~~~&,&~~
\pi_\sigma[k]=\pi_\sigma^N[k]\oplus \pi_\sigma^K[k]\nn\\
\phi^*_\sigma[k]=\phi_\sigma^{*N}[k]\oplus \phi_\sigma^{*K}[k]~~&,&~~
\pi^*_\sigma[k]=\pi_\sigma^{*N}[k]\oplus \pi_\sigma^{*K}[k]~~,
\eea
with $\phi^N,\pi^N,\phi^{*N}, \pi^{*N}\in N_e$ and 
$\phi^K,\pi^K,\phi^{*K}, \pi^{*K}\in K_e$ (and similarly for the antighosts
and their antifields and auxiliary fields).  Note that $\phi^N$ decomposes 
further as $\phi^U\oplus \phi^M$, where $\phi^U$ and 
$\phi^M$ are the exact and coexact pieces; 
similar decompositions hold for all other fields.

\paragraph{Observation}
Due to rank constraints, many extraghost components in a component formalism 
(like the one used for a simple example in the appendix) will generally vanish. 
In this sense, the matrix of extraghosts is sparse; similar remarks apply 
for the auxiliary fields. (Because of this, 
the proper interpretation of various formulae below is to restrict them
to the range of $\sigma$ for which the corresponding fields are non-vanishing. 
To simplify notation, we shall not indicate such ranges explicitly.)
The position of non-vanishing components in the extraghost 
matrix generally corresponds 
to ghost degrees larger than 3 or smaller 
than zero (this is due to the shift in equation (\ref{deg})). 
The reader can easily convince herself that  a general system 
(for example, 
a system containing 3 bundle components $E_n$ of different grades) 
leads to a rather complicated distribution of non-vanishing entries 
and  ghost numbers. The point of the formalism of the present section 
is that it automatically takes such complicated distributions into account, 
for an arbitrary system of bundles $E_n$ -- thereby allowing for a 
general (or `universal') formulation of the gauge-fixing procedure.

\subsection{A gauge-fermion of `delta-function' type}

We next discuss a choice of gauge-fixing fermion
which leads to the gauge used in \cite{gauge}. Due to the presence 
of harmonic modes, our gauge-fixing procedure will be a bit involved, so we 
shall introduce it in two steps. The gauge-fixing fermion will be 
of delta-function (or Landau) type. A more general 
family of gauges (of so-called weighted, or Feynman type) is 
briefly discussed in the appendix; it leads to the same results as the 
Landau-type fermion, after taking an appropriate limit.

\subsubsection{Delta-function gauge-fixing in the absence of zero modes}

Let us first consider the case when the cohomology $H_d^*({\cal H})$ 
is trivial in all degrees. In this situation, one has: 
\be
ker d_e=im d_e~~,~~ker d_e^\dagger =im d_e^\dagger~~, 
~~{\cal H}_e=N_e=im d_e\oplus im d_e^\dagger~~,
\ee 
and the harmonic subspace $K_e$ vanishes. In particular,
$\phi$ coincides with $\phi^N$, and 
similarly for all other fields. 

Following standard gauge-fixing procedure, we consider
the complete action $S_c=S_e+S_{aux}$, where the auxiliary action is
given by:
\bea
\label{Saux_massive}
S_{aux}=Re\int_{L}{str_e\left(\sum_{k\geq 1}\sum_{\sigma\geq 2k}{
\pi^N_\sigma[k]*\phi^{*N}_\sigma[k]}+
\sum_{k\geq 0}\sum_{\sigma\geq 2k+1}{{\overline \pi}^N_\sigma[k]*
{\overline \phi}_\sigma^{*N}[k]}\right)}~~.
\eea
It is easy to check 
that $S_{aux}$ has ghost number zero and is Grassmann-even. 

Consider the following gauge-fixing fermion:
\bea
\label{fermion_delta}
&&\Psi=Re\int_{L}{str_e\sum_{k\geq 0}\left(\sum_{\sigma\geq 2k}{
{\overline \phi}^N_{\sigma+1}[k]*d_ec_e\phi^N_\sigma[k]}+\sum_{\sigma\geq
2k+1}{\phi^N_{\sigma+1}[k+1]* d_ec_e{\overline \phi}^N_\sigma[k]}\right)}~~.
\eea
It is easy
to check that $\Psi$ is Grassmann-odd and
satisfies $gh(\Psi)=-1$.
The gauge-fixed action is obtained in two steps.

\paragraph{Elimination of antifields}

First, one must
eliminate all antifields with the help of the gauge-fixing fermion:
\bea
\label{af_elim}
\phi^{*N}_\sigma[k]=\frac{\delta \Psi}{\delta \phi^N_\sigma[k]}~~,~~
{\overline \phi}^{*N}_\sigma[k]=\frac{\delta \Psi}{\delta {\overline
\phi}^N_\sigma[k]}~~
\eea
as well as: 
\bea
\label{pi_elim}
\pi^{*N}_\sigma[k]={\overline \pi}^{*N}_\sigma[k]=0~~.
\eea
The functional derivatives in the first equation are defined 
as in (\ref{delta}).
Relations (\ref{af_elim}) lead to the equations
\footnote{To arrive at these expressions, we used the identity:
\bea
Re\langle {\hat u}, d_ec_e {\hat v} \rangle_e=
(-1)^{(deg{\hat u}+1)deg {\hat v}}
Re\langle c_ed_e{\hat u}, {\hat v}\rangle_e~~,\nn
\eea
which follows from (\ref{cddc}).}:
\bea
\label{eq_elim}
\phi^{*N}_\sigma[k]&=&c_ed_e{\overline \phi}^N_{\sigma+1}[k]-
d_ec_e{\overline \phi}^N_{\sigma-1}[k-1]~~\nn\\ {\overline
\phi}^{*N}_\sigma[k]&=&c_ed_e\phi^N_{\sigma+1}[k+1]-
d_ec_e\phi^N_{\sigma-1}[k]~~,
\eea
where we defined 
${\overline \phi}^N_\sigma[-1]:=0$. Substituting (\ref{eq_elim})
into $S_c$ gives the gauge-fixed action $S_{gf}=
S_e(\phi^N_\sigma,\phi^{*N}_\sigma=\frac{\delta \Psi}{\delta \phi^N_\sigma})+
S_{aux,gf}$,
where:
\be
\label{Sauxgf_massive}
S_{aux,gf}=S_{aux}\left(\phi^{*N}_\sigma[k]=
\frac{\delta \Psi}{\delta \phi^N_\sigma[k]}~,~{\overline \phi}^{*N}_\sigma[k]=
\frac{\delta \Psi}{\delta {\overline \phi}^N_\sigma[k]}\right)~~.
\ee
According to the general formalism, correlation functions can now be computed 
as:
\be
\label{correlator}
\langle \langle {\cal O}_1\dots {\cal O}_k
\rangle\rangle=\int{\prod_{k,\sigma}{{\cal D}[\phi^N_\sigma[k]]
{\cal D}[{\overline \phi}^N_\sigma[k]]{\cal D}[\pi^N_\sigma[k]] 
{\cal D}[{\overline
\pi}^N_\sigma[k]]}e^{-i\lambda S_{gf}}{\cal O}_1\dots {\cal O}_k}~~.
\ee

\paragraph{Integration of auxiliary fields}

The next step is to notice that the equations of motion for the
auxiliary fields $\pi^N_\sigma[k]$ and ${\overline \pi}^N_\sigma[k]$ impose the
conditions:
{\footnotesize \bea
\label{gf_gen}
&&\phi^{*N}_\sigma[k+1]=
\frac{\delta S_{aux,gf}}{\delta \phi^N_\sigma[k+1]}=0\Leftrightarrow
d_ec_e{\overline \phi}^N_{\sigma-1}[k]=
c_ed_e {\overline
\phi}^N_{\sigma+1}[k+1]~~(k\geq 0~,~\sigma\geq 2k+2)\nn\\
&&{\overline \phi}^{*N}_\sigma[k]=
\frac{\delta S_{aux,gf}}{\delta {\overline \phi}^N_\sigma[k]}=0\Leftrightarrow 
d_ec_e\phi^N_{\sigma-1}[k]=c_ed_e\phi^N_{\sigma+1}[k+1]~~
(k\geq 0~~,~~\sigma\geq 2k+1)~~,
\eea}\noindent
which follow by integrating out $\pi^N_\sigma[k]$ ($k\geq 1$)
and ${\overline \pi}^N_\sigma[k]$ ($k\geq 0$).
Note that the auxiliary action does not contain terms in 
$\pi^N[0]$ (in fact, there are no such auxiliary fields), which is 
why we shifted $k$ by $+1$ in the first equation with respect to 
(\ref{eq_elim}). In particular, the only condition imposed on 
${\overline \phi}^N[0]$ is $d_ec_e{\overline \phi}^N_{\sigma}[0]=
c_ed_e {\overline \phi}^N_{\sigma+2}[1]$. On the other hand, $S_{aux}$ 
{\em does} contain terms in ${\overline \pi}^N[0]$, which explains the range 
of $k$ used in the second equation.

At the path integral level, this can be understood as follows.
Consider a correlator (\ref{correlator}) of
observables ${\cal O}_j$ 
which are independent of the auxiliary fields.  Since the
auxiliary fields appear linearly in the gauge-fixed action, one can
perform the path integral over $\pi^N_\sigma[k]$ and 
${\overline \pi}^N_\sigma[k]$.
Due to the form of $S_{aux,gf}$, this 
produces delta-function factors $\delta(\frac{\delta
\Psi}{\delta \phi_\sigma[k]})$ and $\delta(\frac{\delta \Psi}{\delta
{\overline \phi}_\sigma[k]})$, which restrict the path integral to fields
satisfying the constraints (\ref{gf_gen}).  This effectively
implements (\ref{gf_gen}) at the level of
all correlators, as long as we insist that observables should be
independent of the auxiliary fields. Note that after integrating out 
$\pi$ and ${\overline \pi}$, the term $S_{aux,gf}$ is completely eliminated 
from the gauge-fixed action. 
Thus:
{\footnotesize\bea
\label{correlator2}
\langle \langle {\cal
O}_1\dots {\cal O}_k
\rangle\rangle&=&
\int{\prod_{k,\sigma}{{\cal D}[\phi^N_\sigma[k]]
{\cal D}[{\overline \phi}^N_\sigma[k]]}e^{-i\lambda
S_{e,gf}}{\cal O}_1\dots {\cal O}_k}\\
&&\!\!\!\!\!\!\!\!\!\!\!\!\!\!\!\!
\prod_{k\geq 0}{\prod_{\sigma\geq 2k+1}{\delta(
d_ec_e\phi^N_{\sigma-1}[k]-c_ed_e\phi^N_{\sigma+1}[k+1]
)}}
\prod_{k\geq 0}{\prod_{\sigma\geq 2k+2}{\delta(
d_ec_e{\overline \phi}^N_{\sigma-1}[k]-
c_ed_e {\overline\phi}^N_{\sigma+1}[k+1])}}~~.\nn
\eea}
To eliminate the delta-functions, we decompose $\phi^N$ and 
${\overline\phi}^N$ into their exact and coexact 
components (denoted by superscripts $U$ and $M$, respectively).
Since $d_ec_e$/$c_ed_e$ vanishes on coexact/exact elements, and 
restricts to a bijection on the exact/coexact  subspaces of 
${\cal H}_e$, we can separate the delta-functions as follows: 
{\footnotesize\bea
\label{deltas}
&&\delta(
d_ec_e\phi^N_{\sigma-1}[k]-c_ed_e\phi^N_{\sigma+1}[k+1]
)=\delta(
d_ec_e\phi^U_{\sigma-1}[k])\delta(c_ed_e\phi^M_{\sigma+1}[k+1])=\nn\\
&&~~~~~~~~~~=\left[det'_{\R,s=\sigma-1}(dd^\dagger)
det'_{\R,s=\sigma+1}(d^\dagger d)\right]^{\frac{(-1)^\sigma}{2}}
\delta(\phi^U_{\sigma-1}[k])\delta(\phi^M_{\sigma+1}[k+1])
~~,\nn\\
&&\delta(
d_ec_e{\overline \phi}^N_{\sigma-1}[k]-
c_ed_e {\overline\phi}^N_{\sigma+1}[k+1])=\delta(
d_ec_e{\overline \phi}^U_{\sigma-1}[k])
\delta(c_ed_e {\overline\phi}^M_{\sigma+1}[k+1])=\\
&&~~~~~~~~~~=\left[det'_{\R,s=\sigma-1}
(d d^\dagger)det'_{\R,s=\sigma+1}(d^\dagger d)\right]^{\frac{(-1)^\sigma}{2}} 
\delta(
{\overline \phi}^U_{\sigma-1}[k])
\delta({\overline\phi}^M_{\sigma+1}[k+1])~~.\nn
\eea}\noindent
To derive the last equalities, we noticed that
$\delta(d_ec_e u^U_\sigma)=det'_\R([(dc)_{s=\sigma}]^t
(dc)_{s=\sigma})^{\frac{(-1)^{\sigma+1}}{2}}\delta(u^U_\sigma)$ and 
$\delta(c_ed_e u^M_\sigma)=
det'_\R([(cd)_{s=\sigma}]^t(cd)_{s=\sigma})^{\frac{(-1)^{\sigma+1}}{2}}
\delta(u^M_\sigma)$, for $u^U\in im d_e(s=\sigma)\cap {\cal H}_e^1$ 
and $u^M\in im d_e^\dagger (s=\sigma)\cap {\cal H}_e^1$,
where the prime denotes restriction to the orthocomplement of the relevant 
operator and 
$^t$ denotes the transpose of a real-linear operator with respect to 
the induced Euclidean scalar product $(.,.)=Re <.,.>$ on ${\cal H}$. 
The determinant factors follow from the transformation rule of the 
associated measure, which is controlled by the volume forms constructed 
with the metric $(.,)$; the exponents are controlled by the Grassmannality 
of $u_\sigma$ (which is equal to $\sigma~(mod~2)$). 
To simplify these factors, we used the relations:
\be
(dcu, v)=(-1)^{1+s(u)}(u, cdv)~~,~~(cd u, v)=(-1)^{s(u)}(u, cdv)~~,
\ee
which can be checked by using the relation between the Hermitian product $h$ 
and the bilinear form $<.,.>$, as well as the properties of the later.
These relations imply:
\be
\label{trp}
[(dc)_{s=\sigma}]^t=(-1)^{1+\sigma}(dc)_{s=-2-\sigma}~~,~~[(cd)_{s=\sigma}]^t=
(-1)^\sigma (cd)_{s=-\sigma}~~,
\ee  
which in turn give: 
\be
[(dc)_{s=\sigma}]^t(dc)_{s=\sigma}=dd^\dagger~~,~~
[(cd)_{s=\sigma}]^t(cd)_{s=\sigma}=d^\dagger d~~.
\ee
We note that 
$(dc)_{s=\sigma}$ is a map ${\cal H}(\sigma)\rightarrow {\cal H}(-2-\sigma)$
and $(cd)_{s=\sigma}$ is a map 
${\cal H}(\sigma)\rightarrow {\cal H}(-\sigma)$, while 
their adjoints  $[(dc)_{s=\sigma}]^t$ and  $[(cd)_{s=\sigma}]^t$ give reversed 
arrows between the same subspaces.

Combining everything in equation (\ref{deltas}), we obtain a factor
$PQ$, where:
{\footnotesize \bea
\label{P}
P&:=&\prod_{k\geq 0}{\prod_{\sigma\geq 2k+1}{\delta(\phi^U_{\sigma-1}[k])
\delta(\phi^M_{\sigma+1}[k+1])}}\times 
\prod_{k\geq 0}{\prod_{\sigma\geq 2k+2}{
\delta({\overline \phi}^U_{\sigma-1}[k])
\delta({\overline \phi}^M_{\sigma+1}[k+1])}}=\nn\\
&=&\prod_{\sigma\geq 0}{\delta(\phi^U_{\sigma}[0])}
\prod_{\sigma\geq 1}{\delta({\overline \phi}^U_{\sigma}[0])}\times
\prod_{k\geq 1}{\prod_{\sigma\geq 2k}{
\delta(\phi_\sigma^N[k])}}
\prod_{k\geq 1}{\prod_{\sigma\geq 2k+1}{
\delta({\overline \phi}_\sigma^N[k])}}~~,
\eea}\noindent
while:
{\footnotesize 
\bea
Q&:=&\prod_{k\geq 0}{\prod_{\sigma\geq 2k+1}
{\left[det'_{\R,s=\sigma-1}(dd^\dagger )
det'_{\R,s=\sigma+1}(d^\dagger d)\right]^{\frac{(-1)^\sigma}{2}}}}
\prod_{k\geq 0}{\prod_{\sigma\geq 2k+2}{\left[det'_{\R,s=\sigma-1}
(dd^\dagger )
det'_{\R,s=\sigma+1}(d^\dagger d)\right]^{\frac{(-1)^\sigma}{2}}}}\nn\\
&=&\prod_{k\geq 0}{\prod_{\sigma\geq 2k+1}
{\left[det'_{s=\sigma-1}(dd^\dagger)
det'_{s=\sigma+1}(d^\dagger d)\right]^{(-1)^\sigma}}}
\prod_{k\geq 0}{\prod_{\sigma\geq 2k+2}{\left[det'_{s=\sigma-1}(dd^\dagger)
det'_{s=\sigma+1}(d^\dagger d)\right]^{(-1)^\sigma}}}~~,
\eea}\noindent
where in the second line we passed to complex determinants. 
To simplify $Q$, we use the fact that the operators 
$dd^\dagger|_{ker (d^\dagger)^\perp(s=\sigma)}$ and 
$d^\dagger d|_{ker(d)^\perp(s=\sigma+1)}$ are isospectral (this is 
proved in \cite{rs}), which implies that their determinants are equal. 
Thus: 
{\footnotesize \be
Q=\prod_{k\geq 0}{\prod_{\sigma\geq 2k+1}
{\left[det'_{s=\sigma}(d^\dagger d)
det'_{s=\sigma+1}(d^\dagger d)\right]^{(-1)^\sigma}}}
\prod_{k\geq 0}{\prod_{\sigma\geq 2k+2}{\left[det'_{s=\sigma}(d^\dagger d)
det'_{s=\sigma+1}(d^\dagger d)\right]^{(-1)^\sigma}}}=\frac{1}{I}~~,
\ee} 
where: 
\be
\label{I}
I:=\prod_{\sigma>0}{det'_{s=\sigma}(d^\dagger d)^{(-1)^{\sigma+1}}}~~.
\ee

\paragraph{The gauge-fixed path  integral}

Returning to expression (\ref{correlator2}), the factor (\ref{P}) 
eliminates all integrals over $\phi_\sigma^N[k]$ and 
${\overline \phi}_\sigma^N[k]$
(with $k\geq 1$) 
by setting these fields to zero, and restricts the remaining integrals 
to integrals over the {\em coexact} components $\phi^M[0]=\phi^M$ and 
${\overline \phi}^M[0]:={\overline \phi}^M$. On the other hand, 
$Q$ produces the prefactor $\frac{1}{I}$. Since the contribution $S_{aux}$ 
was eliminated when integrating out the auxiliary fields, the only other 
effect of gauge-fixing is to impose the condition 
$\phi^*_\sigma=c_ed_e{\overline \phi}^M_{\sigma+1}$ 
in the extended action $S_e$.
Remembering that $\phi^*_\sigma={\hat \phi}_{-1-\sigma}$,  
this also reads:
\be
{\hat \phi}_{-\sigma}=c_ed_e{\overline \phi}^M_\sigma~~
{\rm~for~all~}~~\sigma>0~~.
\ee
Recalling the notations
${\hat \phi}_+:=\oplus_{\sigma\geq 0}{\phi_\sigma}$,  
${\hat \phi}_-:=\oplus_{\sigma<0}{\phi_\sigma}$ and 
${\overline \phi}:=\oplus_{\sigma>0}{{\overline \phi}_\sigma}$,
we have ${\hat \phi}_-=c_ed_e{\overline \phi}$. Note that $c_ed_e$ gives a 
bijection between $im d_e^\dagger (s>0)$ and $im d_e^\dagger (s<0)$,
so the last relation can be viewed as a (differential) linear change 
of variables.

Thus the gauge-fixed action reduces to:
\be
\label{Sgf}
S_{gf}=S_e(\phi^M_\sigma, \phi^{*M}_\sigma=c_ed_e {\overline \phi}^M_{\sigma+1})=
S_e({\hat \phi}^M_+, {\hat \phi}^M_-=c_ed_e{\overline \phi}^M)~~,
\ee
where ${\hat \phi}^M_+\in im d_e^\dagger (s\geq 0)$ and 
${\overline \phi}^M\in im d_e^\dagger (s>0)$. 
The gauge-fixed, extended field ${\hat \phi}^M={\hat \phi}_+^M
\oplus {\hat \phi}^M_-=
{\hat \phi}^M_+\oplus (c_ed_e {\overline \phi}^M)$ has the expansion:
\be
{\hat \phi}^M=\dots +c_ed_e{\overline \phi}^M_2+
c_ed_e{\overline \phi}^M_1+\phi_0^M+\phi_1^M+\phi_2^M+\dots~~.
\ee
Thus the sole effect of gauge-fixing is to replace the extended field
${\hat \phi}={\hat \phi}_+\oplus {\hat \phi}_- \in {\cal H}_e$ with the 
field ${\hat \phi}^M_+\oplus (c_ed_e{\overline\phi}^M)\in im d_e^\dagger$. 
This is very similar to what happens 
for acyclic backgrounds in the ungraded case \cite{AS1}.

\paragraph{Conclusion}

In the acyclic case, correlators have the form:
\be
\label{correlator_acyclic}
\langle \langle {\cal O}_1\dots {\cal O}_k\rangle\rangle=
\frac{1}{I}\int{\prod_{\sigma\geq 0}{{\cal D}[\phi^M_\sigma]}
\prod_{\sigma\geq 1}{{\cal D}[{\overline \phi}^M_\sigma]}
e^{-i\lambda S_e(\phi^*_\sigma=c_ed_e{\overline \phi}^M_{\sigma+1}~,~ 
\phi_\sigma=\phi^M_\sigma)}{\cal O}_1\dots {\cal O}_k}~~,
\ee
where $I$ is given in equation (\ref{I}).

\subsubsection{Extension to the case with zero modes}

According to Hodge theory, the field space now splits into
\be
{\cal H}_e=N_e\oplus K_e
\ee
where $N_e=im d_e\oplus im d_e^\dagger$ 
is, as before, the collection of all exact and coexact field configurations
while $K_e$ is the collection of all harmonic modes. 
Accordingly, we write ${\hat \phi}={\hat \phi}^N\oplus {\hat \phi}^K$
(with ${\hat \phi}^N\in N_e$ and ${\hat \phi}^K\in K_e$)
and perform gauge fixing for each component 
separately, by adding trivial pairs 
along each subspace. The full auxiliary action reads:
{\footnotesize \bea 
S_{aux}&=&Re\int_{L}{str_e\left(\sum_{k\geq 1}\sum_{\sigma\geq 2k}{ 
\pi^N_\sigma[k]*\phi^{*N}_\sigma[k]}+ 
\sum_{k\geq 0}\sum_{\sigma\geq 2k+1}{{\overline \pi}^N_\sigma[k]* 
{\overline \phi}^{*N}_\sigma[k]}\right)}\\ 
&{}&\!\!\!\!\!\!\!\!
+Re\int_{L}{str_e\left(\sum_{k\geq 1}\sum_{\sigma\geq 2k}{ 
\pi^K_\sigma[k]*A_{-1-\sigma}[k]^{1/2}\phi^{*K}_\sigma[k]}+ 
\sum_{k\geq 0}\sum_{\sigma\geq 2k+1}{ 
{\overline \pi}^K_\sigma[k]*A_{-1-\sigma}[k]^{1/2} 
{\overline \phi}^{*K}_\sigma[k]}\right)}\nonumber\\ 
&{}&\!\!\!\!\!\!\!\!
+Re\int_{L}{str_e\sum_{k\geq 0}\left(\sum_{\sigma\geq 2k}{ 
\lambda^K_\sigma[k]*A_{-1-\sigma}[k]^{1/2}{\overline\mu}^{*K}_\sigma[k]}+ 
\sum_{\sigma\geq 2k+1} 
{{\overline \lambda}^K_\sigma[k]* A_{-1-\sigma}[k]^{1/2}
{\mu}^{*K}_\sigma[k]}\right)}~~, 
\nonumber
\eea}\noindent  
where $A_\sigma[k]$ are some (strictly) 
positive operators on $K(\sigma)$, while 
the new fields $\lambda^K$, $\mu^K$ , ${\overline \lambda}^K$ and 
${\overline \mu}^K$ have charges:
{\footnotesize \bea
\label{charges}
s(\lambda_\sigma^K[k])=s({\overline \lambda}_\sigma^K[k])=\sigma~&,&~
s(\mu_\sigma^K[k])=s({\overline \mu}_\sigma^K[k])=\sigma\nn\\
g(\lambda_\sigma^K[k])=g({\overline \lambda}_\sigma^K[k])=\sigma ~(mod~2)~&,&~
g(\mu_\sigma^K[k])=g({\overline \mu}_\sigma^K[k])=\sigma+1 ~(mod~2)\nn\\
gh(\lambda_\sigma^K[k])=2k-\sigma~&,&~
gh(\mu_\sigma^K[k])=\sigma-2k-1\nn\\
gh({\overline \lambda}_\sigma^K[k])=
\sigma -2k~&,&~
gh({\overline \mu}_\sigma^K[k])=2k-\sigma-1~~.
\eea}\noindent
Note that  $\phi^{K*}_\sigma$ etc have $s=-1-\sigma$. 
As in \cite{rs}, we shall choose $A_\sigma[k]$ such that: 
\be
\label{cA}
cA_\sigma[k]=A_{-1-\sigma}[k]c~~.
\ee
This is possible because $c$ gives bijections between $K(\sigma)$ and 
$K(-1-\sigma)$. 

The charge assignments in (\ref{charges}) follow from elementary 
considerations. For example, the BRST variation 
of ${\overline\mu}^K_\sigma[k]$ is $\lambda^K_\sigma[k]$, which implies that
their Grassmann parities should be opposite, their $U(1)$ degrees should
be the same (and consequently their $s$-numbers) while the ghost number
of $\lambda^K_\sigma[k]$ should be larger by one unit than the ghost number
of ${\overline\mu}^K_\sigma[k]$. Similar arguments hold for ${\mu}^K_\sigma[k]$
and ${\overline\lambda}^K_\sigma[k]$. The terms involving
$\lambda$ and ${\overline\lambda}$ will allow us 
to gauge-fix 
the harmonic components of all fields $\phi$ and ${\overline\phi}$.
The range of summation of $\sigma$ and $k$ implies that
$\lambda$ will fix the harmonic components of $\phi$ while
${\overline\lambda}$ will fix the harmonic components of ${\overline\phi}$.
Thus, the quantum numbers of $\lambda$ and ${\overline\lambda}$
are uniquely fixed which in turn fixes the rest.

The simplest gauge is obtained by choosing a delta-type gauge fixing  fermion
for both harmonic and non-harmonic components. This has the form:
{\footnotesize\bea 
\Psi&=&Re\int_{L}{str_e\sum_{k\geq 0} 
\left( \sum_{\sigma\geq 2k}{ 
{\overline \phi}^N_{\sigma+1}[k]*d_ec_e\phi^N_\sigma[k]}+ 
\sum_{\sigma\geq 2k+1} 
{\phi^N_{\sigma+1}[k+1]* d_ec_e{\overline \phi}^N_\sigma[k]}\right)} 
\\ 
&+&
Re\int_{L}str_e\sum_{k\geq 0}\left( \sum_{\sigma\geq 2k}{ 
(c_e{\overline\mu}^K_{\sigma}[k])*A_\sigma[k]^{1/2}
(\phi^K_\sigma[k]-\phi^H_\sigma[k])}\right.\nn\\
&&~~~~~~~~~~~~+ \left.
\sum_{\sigma\geq 2k+1} 
{(c_e\mu^K_{\sigma}[k])* A_\sigma[k]^{1/2} 
({\overline \phi}^K_\sigma[k]-{\overline \phi}^H_\sigma[k])}\right)\nonumber~~,
\eea}\noindent where $\phi^H_\sigma[k]$ and ${\overline \phi}^H_\sigma[k]$ 
are some 
constant elements belonging to the appropriate harmonic subspaces
(these constant shifts serve as parameters). 
In principle, for the terms in the first line above one could 
forget the superscript $N$ and consider the full field, because 
harmonic components are annihilated by $d_e$ and $d_ec_e$. These
terms would then be invariant under shifting $\phi$ and
${\overline\phi}$ by harmonic forms. 
This residual gauge symmetry is then fixed by the terms 
in the second line. If one did not fix this extra invariance then the
antifields would not be uniquely determined in terms of fields and
 the resulting kinetic term would still be degenerate.

The harmonic components of antifields are eliminated through the equations: 
{\footnotesize \bea 
\phi^K_\sigma{}^*[k]&=&\frac{\delta \Psi}{\delta \phi^K_\sigma[k]} 
=c_eA_\sigma[k]^{1/2}{\overline\mu}^K_\sigma[k]~~~~~~~~~~,~~~~~~~~~
{\overline \phi}^K_\sigma{}^*[k]= 
\frac{\delta \Psi}{\delta {\overline\phi}^K_\sigma[k]} 
=c_eA_\sigma[k]^{1/2}{\mu}^K_\sigma[k]\\ 
\mu^K_\sigma{}^*[k]&=&\frac{\delta \Psi}{\delta \mu^K_\sigma[k]} 
=c_eA_\sigma[k]^{1/2}
({\overline\phi}_\sigma^K[k]- {\overline\phi}_\sigma^H[k])~~,~~
{\overline\mu}^K_\sigma{}^*[k]=\frac{\delta \Psi}{\delta  
{\overline\mu}^K_\sigma[k]}=c_eA_\sigma[k]^{1/2}
(\phi_\sigma^K[k]-\phi_\sigma^H[k]) ~~,\nn 
\eea} 
while the non-harmonic components are eliminated through (\ref{eq_elim}). 

Substituting these into the auxiliary action gives:
{\footnotesize \bea 
\label{sgf}
S_{aux, gf}&=& 
Re\int_{L}str_e\left(\sum_{k\geq 0}\sum_{\sigma\geq 2k}
\lambda^K_\sigma[k]* 
c_eA_\sigma[k]({\phi}_\sigma^K[k]- {\phi}_\sigma^H[k])
\right.\nonumber\\
&{}&\left.~~~~~~~~~~~~~~~~+ 
\sum_{k\geq 0}\sum_{\sigma\geq 2k+1} 
{\overline \lambda}^K_\sigma[k]* 
c_eA_{\sigma}[k]({\overline\phi}_\sigma^K[k]-{\overline\phi}_\sigma^H[k])
\right) \nonumber
\\ 
&+&Re\int_{L}str_e\left(\sum_{k\geq 1}\sum_{\sigma\geq 2k}{ 
\pi^K_\sigma[k]*c_eA_\sigma[k]{\overline\mu} 
_\sigma^K[k]}\right.\nonumber\\
&{}&\left.~~~~~~~~~~~~~~~~+\sum_{k\geq 0}\sum_{\sigma\geq 2k+1}{ 
{\overline \pi}^K_\sigma[k]*c_e 
A_\sigma[k]\mu^K_\sigma[k]}\right) \nonumber\\
&+&S_{aux,gf}(\phi^N, {\overline\phi}^N)~~,
\eea} 
where the last term has the form given in 
(\ref{Sauxgf_massive},\ref{Saux_massive}) and where we used relation 
(\ref{cA}).

Considering a correlator of operators 
independent of auxiliary fields, the 
path integral over $\lambda^K$ and ${\overline\lambda}^K$
produces:
{\footnotesize \bea
&&\prod_{k\ge 0;\sigma\geq 2k }{\delta(\phi^K_\sigma[k] -\phi^H_\sigma[k] )}
\prod_{k\ge 0;\sigma\geq 2k+1 }{\delta({\overline\phi}^K_\sigma[k] - 
{\overline\phi}^H_\sigma[k])}\times\\
&&\times\prod_{k\ge 0; \sigma\ge 2k}
det_\R A_\sigma[k]^{(-1)^{\sigma+1}}
\prod_{k\ge 0; \sigma\ge 2k+1}
det_\R A_\sigma[k]^{(-1)^{\sigma+1}}\nn
\label{li}~~,
\eea}\noindent
where we used the fact that the Grassmannality of $\phi^K_\sigma[k]$ and 
${\overline\phi}^K_\sigma[k]$ equals $\sigma~(mod~2)$.

On the other hand, 
the integral over $\pi^K$, $\mu^K$ and their barred counterparts 
gives:
{\footnotesize \bea
\label{mupi}
\prod_{k\ge 1; \sigma\ge 2k}
det_\R A_\sigma[k]^{(-1)^{\sigma}}
\prod_{k\ge 0; \sigma\ge 2k+1}
det_\R A_\sigma[k]^{(-1)^{\sigma}}
\label{rest}~~.
\eea}\noindent 
To arrive at this equation, we first performed the integral over $\mu^K$ and 
${\overline\mu}^K$, which gives delta-function factors of the form
$\delta(A_\sigma[k]\pi^K_\sigma[k])$
and $\delta(A_\sigma[k]{\overline \pi}^K_\sigma[k])$, 
and we noticed that: 
\bea
\delta(A_\sigma[k]\pi^K_\sigma[k])&=&
det_\R A_\sigma[k]^{(-1)^\sigma}
\delta(\pi^K_\sigma[k])~~\nn\\
\delta(A_\sigma[k]\pi^K_\sigma[k])&=&
det_\R A_\sigma[k]^{(-1)^\sigma}
\delta({\overline \pi}^K_\sigma[k])~~,
\eea
where we used the fact that the Grassmannality of $\pi^K_\sigma[k]$ 
and ${\overline \pi}^K_\sigma[k]$ is $\sigma+1~(mod~2)$. The 
remaining integral over these fields eliminates the delta-function factors 
and gives the contribution (\ref{mupi}).

Therefore, the final result of the integral over the harmonic sector of
auxiliary fields is to produce a factor:
{\footnotesize \bea
\label{harmonic_factor}
\prod_{k\ge 0;\sigma\geq 2k}{\delta(\phi^K_\sigma[k] -\phi^H_\sigma[k])}
\prod_{k\ge 0;\sigma\geq 2k+1}{\delta({\overline\phi}^K_\sigma[k] -
{\overline\phi}^H_\sigma[k])}
\prod_{\sigma\ge 0}
[det_\R A_\sigma]^{(-1)^{\sigma+1}}~~,
\eea}\noindent
while eliminating all but the last term in eq. (\ref{sgf}).
In the last relation, we defined $A_\sigma:=A_\sigma[0]$.
The contribution (\ref{harmonic_factor}) 
has the effect of killing all integrals over all harmonic 
components, while inducing a prefactor: 
\be
J:=\prod_{\sigma\ge 0}
det_\R A_\sigma ^{(-1)^{\sigma+1}}
\ee 
in the gauge-fixed path integral. 
For $k=0$, (\ref{harmonic_factor}) gives 
the gauge-fixing condition for  massless modes:
\be
\label{gf_harm}
\phi^K_\sigma=\phi^H_\sigma~~,~~
{\overline \phi}^K_\sigma={\overline\phi}^H_\sigma~~,
\ee
where we defined $\phi^H_\sigma:=\phi^H_\sigma[0]$ and 
${\overline\phi}^H_\sigma:={\overline \phi}^H_\sigma[0]$. 
Since $\phi^K_\sigma$ are 
frozen to the values $\phi^H_\sigma$, 
we are left with the extended action 
$S_e(\phi^*_\sigma, \phi_\sigma=\phi_\sigma^N\oplus \phi^H_\sigma)$ and
with the 
last term of the auxiliary action 
(\ref{sgf}), which depends only on non-harmonic modes.
Integrating out the non-harmonic auxiliary fields $\pi^K$ and 
${\overline \pi}^K$ now has the effect discussed in the previous subsection, 
thereby eliminating the last term in the auxiliary action 
and producing the prefactor $\frac{1}{I}$ with $I$ given in 
equation (\ref{I}). It also 
kills all integrals over extraghosts and their antighosts, 
restricts the integrals over ghosts and antighosts to their massive components 
and  
implements the condition $\phi^*_\sigma=c_ed_e{\overline\phi}^M_{\sigma+1}$ 
in the tree-level BV action $S_e$. The result is a gauge-fixed action 
given by $S_e(\phi^*_\sigma=c_ed_e{\overline \phi}^M_\sigma, \phi_\sigma=
\phi^M_\sigma\oplus \phi^H_\sigma)$, and 
a prefactor equal to $J/I$ in front of the gauge-fixed path integral.

\subsection{Summary}

Combining everything, we find that the result of gauge fixing is as follows:

\

\noindent (1)The gauge-fixed action is given by:

\be
S_{gf}=
S_e(\phi^{*}_\sigma=c_ed_e{\overline\phi}^M_{\sigma+1}, 
\phi_\sigma=\phi^M_\sigma\oplus \phi^H_\sigma)~~,
\ee
where the physical field and ghosts $\phi^M_\sigma~(\sigma \geq 0)$ and 
the antighosts ${\overline \phi}^M_{\sigma}~(\sigma>0)$ belong to 
$im d_e^\dagger$, while $\phi^H_\sigma$ ($\sigma\geq 0$) are some 
fixed harmonic elements which play the role of parameters. 
The components $\phi^H_\sigma$ can be assembled 
into the element:
\be
{\hat \phi}^H_+=\oplus_{\sigma\geq 0}{\phi^H_\sigma}\in K_e(s\geq 0)~~.
\ee
The harmonic components of the antighosts are fixed to some irrelevant 
values, and do not enter the gauge-fixed action.

We also write:
\be
S_{gf}=S_e({\hat \phi}_-=c_ed_e{\overline  \phi}^M, 
{\hat \phi}_+={\hat \phi}^M_+\oplus {\hat \phi}^H_+)~~,
\ee
where ${\overline\phi}^M=\sum_{\sigma >0}{{\overline \phi}^M_\sigma}\in 
im d_e^\dagger(s>0)$. Note that ${\hat \phi}_+=\sum_{\sigma\geq 0}
{{\hat \phi}_\sigma}\in im d_e^\dagger(s\geq 0)\oplus 
K_e(\sigma \geq 0)$ and 
 ${\hat \phi}_-=\sum_{\sigma<0}
{{\hat \phi}_\sigma}\in im d_e^\dagger(s<0)$.

\

\noindent (2) A prefactor of  $J/I$, where:
\be
\label{J}
J=
\prod_{\sigma\ge 0}
det_\R A_\sigma ^{(-1)^{\sigma+1}}~~{\rm and}~~
I=\prod_{\sigma>0}{det'_{s=\sigma}(d^\dagger d)^{(-1)^{\sigma+1}}}~~,
\ee\noindent
is induced in the path integral measure. 
The data 
$A_\sigma:=A_\sigma[0]$ are (strictly) positive operators on the 
harmonic subspaces $K(\sigma)$, which play the role of gauge-fixing 
parameters for the zero modes.

\subsubsection{Cohomological formalism for $A_\sigma$}

To make contact with the formalism of \cite{rs}, we now express the 
data $A_\sigma$ in terms of equivalent data used in that paper. 
Let us fix $\sigma$ and consider the Hermitian metric $h_\sigma$ induced
by $h$ on $K_\sigma$. It is easy to see that specifying a positive operator 
$A_\sigma$ on $K_\sigma$ is equivalent to specifying another Hermitian 
metric $g_\sigma$ on $K_\sigma$. Indeed, given such an operator 
one constructs $g_\sigma$ through:
\be
\label{g_sigma}
g_\sigma(u,v):=h_\sigma((A_\sigma)^2 u, v)=h_\sigma(A_\sigma u, A_\sigma v)~~.
\ee
Conversely, any Hermitian metric can be written in this form.
To see this, consider bases $e_i$ and $e'_i$ of $K_\sigma$ which are 
orthonormal with respect to $h_\sigma$, respectively $g_\sigma$. 
If $B_\sigma$ is the invertible operator which takes $e'_i$ into $e_i$, 
then we have:
\be
h_\sigma(B_\sigma(e'_i), B_\sigma(e'_j))=h_\sigma(e_i,e_j)=g_\sigma(e'_i,e'_j)=
\delta_{ij}\Rightarrow g_\sigma(u,v)=h_\sigma(B_\sigma u, B_\sigma v)~~.
\ee
Defining the positive operator $A_\sigma$ through 
$A_\sigma:=(B_\sigma^\dagger B_\sigma)^{1/2}$, we have $B_\sigma^\dagger 
B_\sigma=(A_\sigma)^2$ and $g_\sigma(u, v)=h_\sigma(B_\sigma u, B_\sigma v)=
h_\sigma((A_\sigma)^2u, v)$.

We next consider the Hodge isomorphism $f_\sigma:K_\sigma\rightarrow 
H_\sigma({\cal H})=H^{1-\sigma}_d({\cal H})$, 
given by associating to every harmonic element $u$ its cohomology class.
It is clear that specifying a Hermitian metric $g_\sigma$ on $K_\sigma$ 
is equivalent to giving a Hermitian metric $h^H_\sigma$ on 
$H_\sigma({\cal H})$, the two being related by:
\be
h_\sigma^H(f_\sigma(u), f_\sigma(v))=g_\sigma(u,v)~~.
\ee
Combining these two observations, we see that the data 
$A_\sigma$ is equivalent to the specification of a metric $h_\sigma^H$ 
on $H_\sigma({\cal H})$, the relation between these equivalent objects being 
given by:
\be
\label{adjoints}
h^H_\sigma(f_\sigma u, f_\sigma v)=h_\sigma((A_\sigma)^2 u, v)~~.
\ee
If $f_\sigma^\dagger : H_\sigma({\cal H})\rightarrow K_\sigma$ 
is the Hermitian conjugate of $f_\sigma$ with respect to the metrics 
$h_\sigma$ and $h^H_\sigma$, then $h^H_\sigma(f_\sigma u, f_\sigma v)=
h_\sigma(f_\sigma^\dagger f_\sigma u, v)$. Combined with 
(\ref{adjoints}), this gives:
\be
f_\sigma^\dagger f_\sigma=(A_\sigma)^2\Leftrightarrow 
A_\sigma=(f_\sigma^\dagger f_\sigma)^{1/2}~~.
\ee
Hence the factor $J=\prod_{\sigma\ge 0}
det_\R A_\sigma ^{(-1)^{\sigma+1}}=\prod_{\sigma\ge 0}
det(A_\sigma)^{2(-1)^{\sigma+1}}$ can also be written as:
\be
\label{Jf}
J=\prod_{\sigma\ge 0}
det(f^\dagger_\sigma f_\sigma)^{(-1)^{\sigma+1}}~~,
\ee
which recovers the prefactor obtained in \cite{rs}. 

As a last observation, we recall that $A_\sigma$ are chosen to satisfy 
relation (\ref{cA}). Via relation (\ref{g_sigma}), the metrics they determine 
on $K(\sigma)$ will obey:
\be
\label{gc}
g_\sigma(cu, cv)=g_{-1-\sigma}(v, u)~~, 
\ee
where we used the fact that $c$ is anti-unitary:
\be
h_\sigma(cu, cv)=h_{-1-\sigma}(v, u)~~
\ee
(this follows from the last of properties (\ref{c_props})). 
As in \cite{rs}, we introduce antilinear operators 
$c_{*,\sigma}$ from $H_\sigma({\cal H})$ to 
$H_{-1-\sigma}({\cal H})$ through the conditions:
\be
c_{*,\sigma} f_\sigma=f_{-1-\sigma} c_\sigma~~.
\ee
The relations (\ref{adjoints}) and (\ref{gc}) imply:
\be
h^H_{-1-\sigma}(c_{*,\sigma}u, c_{*,\sigma}v)=h^H_\sigma(v, u)~~,
\ee
thereby recovering the constraint on $h_\sigma^H$ which was used in \cite{rs}. 
In the BV approach, this condition (which is equivalent with (\ref{cA})), 
is used to bring the gauge-fixed auxiliary action to the form (\ref{sgf}). 

\subsection{Simplified form of the gauge-fixed action}

Let us consider a correlator of the form:
\be
\label{corr}
\langle {\cal O}_0...{\cal O}_n\rangle=\frac{J}{I}
\int{{\cal D}[{\hat \phi}^M_+]{\cal D}[{\overline \phi}^M]e^{-i\lambda
S_e({\hat \phi}_-=c_ed_e{\overline \phi}^M, {\hat \phi}_+=
{\hat \phi}^M_+\oplus {\hat \phi}^H_+)}
{\cal O}_1\dots {\cal O}_n}~~,
\ee
where the observables ${\cal O}$ depend only on the physical field. 
Since $c_ed_e$ is invertible on the subspace $im d_e^\dagger$, one
can perform the change of variables 
${\hat \phi}^M_-=c_ed_e{\overline \phi}^M$. 
This is a linear transformation, under which the path integral 
measure transforms as:
\be
{\cal D}[{\overline \phi}^M]=
R {\cal D}[{\hat \phi}^M_-]~~.
\ee
The prefactor $R$ 
is clearly independent of all fields (though it does depend on the background 
superconnection and metric data). 
This allows us to replace the integral over antighosts 
with an integral over ${\hat \phi}^M_-\in im d_e^\dagger(s<0)$, 
up a modification of the normalization factor:
\be
\label{transformed}
\langle {\cal O}_0...{\cal O}_n\rangle=
J\frac{R}{I}\int{\prod_{{\hat \phi}^M\in imd_e^\dagger}{
e^{-i\lambda S_e({\hat \phi}^M\oplus {\hat \phi}^H_+)}
{\cal O}_1\dots {\cal O}_n}}~~.
\ee
Thus we can ignore the 
substitution $\phi^*=c_ed_e{\overline \phi}$ in the 
gauge-fixed action, and simply replace it 
with the restriction of $S_e$ to the subspace $im d_e^\dagger$, up to 
an appropriate shift by ${\hat \phi}^H_+$.  

The quantity $R$ describes the change of the measure under the transformation
$(c_ed_e)^{-1}:{\cal H}_e(s<0)\rightarrow {\cal H}_e(s>0)$. 
To determine this, we view ${\cal H}$ as a real vector space upon restriction
of the field of scalars. In this case, one has:
\be
R=\prod_{\sigma >0}{det_\R((cd)^t(cd))^\frac{(-1)^{\sigma+1}}{2}}~~,
\ee
where $det_\R$ denotes the determinant of its argument viewed as a 
real-linear map and $(cd)^t$ is the adjoint of the real-linear operator $cd$ 
with respect to the induced Euclidean scalar product $(.,.)=Re h(.,.)$.
Since $cd$ is selfadjoint with respect to this product, 
we have $(cd)^t(cd)=cdcd=d^\dagger d$. Thus:
\be
\label{R}
R=\prod_{\sigma >0}{det_\R(d^\dagger d)^\frac{(-1)^{\sigma+1}}{2}}
=\prod_{\sigma >0}{det(d^\dagger d)^{(-1)^{\sigma+1}}}=I~~,
\ee
where $det$ stands for the complex determinant and $I$ is the quantity 
defined in (\ref{I}). In particular, expression (\ref{transformed}) becomes:
\be
\label{transformed2}
\langle {\cal O}_0...{\cal O}_n\rangle=
J\int{\prod_{{\hat \phi}^M\in imd_e^\dagger}{
e^{-i\lambda S_e({\hat \phi}^M\oplus {\hat \phi}^H_+)}
{\cal O}_1\dots {\cal O}_n}}~~.
\ee
The formalism encoded by this relation is similar to the 
description of (ungraded) gauge-fixed Chern-Simons theory used in 
\cite{AS1}. It has the advantage that its only reference to metric data 
enters through the particular decomposition of the space of extended fields 
into subspaces of harmonic, exact and coexact configurations. On the other
hand, this description treats the physical field and its ghost/antighost 
counterparts in a unified manner. This leads to a simple form of the 
perturbation expansion, as we shall see in Section 6. We stress that this 
formalism and that described by (\ref{corr}) are completely 
equivalent, being related by a change of variables in the path integral.

\section{The partition function and semiclassical approximation}

Let us consider the path integral ($\lambda$ is the
dimensionless coupling constant):
\bea
\label{PI}
Z({\hat \phi}^H)=\frac{J}{I} \int{{\cal D}[{\hat \phi}^M_+, {\overline \phi}^M]
e^{-i\lambda S_{gf}}}=
J\int
{{\cal D}[{\hat \phi}^M]
e^{-i\lambda S_e({\hat \phi}^M\oplus {\hat \phi}^H)}}~~{\rm~for~}~~
{\hat \phi}^H\in K_e~~.
\eea
In the case when ${\hat \phi}^H=\phi^H$ (i.e. the harmonic 
components of ghosts and antighosts are fixed to zero), this 
quantity can be viewed as the partition function in the background 
which results from original superconnection by shifting through $\phi^H$.  
In this section, we study the semiclassical (i.e. Gaussian) approximation 
of (\ref{PI}). 

\subsection{The semiclassical approximation}

The semiclassical approximation to $Z({\hat \phi}^H)$ results upon neglecting 
cubic terms; this amounts to keeping only the kinetic (quadratic) terms 
of the gauge-fixed action. 
Since the kinetic term is independent of ${\hat \phi}^H$,
the resulting quantity $Z_{scl}$ is also ${\hat \phi}^H$-independent
and can be computed by setting ${\hat \phi}^H=0$.
It is instructive to perform the computation in two ways. 

\subsubsection{The first approach} 

Let us start from the equation:
\be
\label{Zf1}
Z_{scl}=\frac{J}{I}\int{
{\cal D}[\prod_{\sigma >0}{\phi^M_\sigma}]
{\cal D}[\prod_{\sigma >0}{{\overline \phi}^M_\sigma}]e^{-i\lambda
S_{gf,kin}}}~~,
\ee
where:
\be
S_{gf,kin}=Re\left[\langle \phi^M_0,d_e\phi^M_0\rangle_e + 
\sum_{\sigma>0}
{\left(\langle c_ed_e{\overline \phi}^M_\sigma, d_e\phi^M_\sigma\rangle_e+
\langle \phi^M_\sigma, d_ec_ed_e{\overline \phi}^M_\sigma\rangle_e
\right)}\right]~~.
\ee

To perform this Gaussian integral, we consider the scalar product
$({\hat u}, {\hat v})_e:=Re h_e({\hat u}, {\hat v})$ 
on ${\cal H}_e$, where ${\cal H}$ is viewed as a real vector space 
by restriction of scalars. This allows us to write:
\bea
S_{gf, kin}&=&(\phi_0, c_ed_e\phi_0)_e
+2\sum_{\sigma>0}{(-1)^\sigma ({\overline \phi}^M_\sigma, 
d_e^\dagger d_e\phi^M_\sigma)_e}~~.
\eea

This expression 
involves the restriction of the Hermitian operator $d^\dagger d$ to the 
subspaces $V_\sigma:=im d^\dagger\cap {\cal H}(\sigma)=(kerd)^\perp
\cap {\cal H}(\sigma)$ (with $\sigma >0)$ as well as the 
restriction of $cd$ to $V_0$. 
Viewing these as real-linear selfadjoint operators, we obtain:
\be
Z_{scl}=ct \times \frac{J}{I}
det'_{\R,s=0}(cd)^{-1/2}\prod_{\sigma>0}{det'_{\R, s=\sigma}
(d^\dagger d)^{(-1)^{\sigma+1}}}~~,
\ee
where $det'_{\R,s=\sigma}(O)$ stands for the real determinant 
of an operator $O$ defined on
${\cal H}(\sigma)$, after its restriction to the orthogonal complement of 
its kernel. 
Using $det'_{\R, s=\sigma}
(d^\dagger d)=det'_{s=\sigma}(d^\dagger d)^2$ and 
partially regularizing $det'_{\R,s=0}(cd)$ by replacing the indefinite 
operator $cd_{s=0}$ with $|cd_{s=0}|=
\sqrt{(cd)^2_{s=0}}=(d^\dagger d_{s=0})^{1/2}$, 
we obtain:
\be
\label{Zscl}
Z_{scl}=ct \times J
det'_{s=0}(d^\dagger d)^{-1/2}
\prod_{\sigma>0}{det'_{s=\sigma}
(d^\dagger d)^{(-1)^{\sigma+1}}}~~,
\ee
where we used expression (\ref{I}) for $I$. 
The result (\ref{Zscl}) can be regularized as explained below.

\subsubsection{The second approach}
Expression (\ref{Zscl}) can also be obtained by starting with the equation:
\be
\label{Zf12}
Z_{scl}=J\int_{{\hat \phi}^M\in im d_e^\dagger}{
{\cal D}[{\hat \phi}^M]e^{-i\lambda S_{e,kin}({\hat \phi}^M)}}
\ee
where:
\bea
S_{e,kin}({\hat \phi}^M)&=&Re\left[
\langle \phi^M_0,d_e\phi^M_0\rangle_e + \sum_{\sigma>0}
{\left(\langle {\hat \phi}^M_{-\sigma}, d_e{\hat \phi}^M_\sigma\rangle_e+
\langle d_e{\hat \phi}^M_{-\sigma}, {\hat \phi}^M_\sigma \rangle_e
\right)}\right]=\nn\\
&=&(\phi^M_0,c_ed_e\phi^M_0)_e + 2\sum_{\sigma>0}{(-1)^\sigma
({\hat \phi}^M_{-\sigma}, c_ed_e{\hat \phi}^M_\sigma)_e}~~.
\eea
This gives:
\bea
Z_{scl}&=&ct \times J
det'_{\R,s=0}(cd)^{-1/2}\prod_{\sigma>0}{det'_{\R, s=\sigma}
(d^\dagger d)^{\frac{(-1)^{\sigma+1}}{2}}}=\nn\\
&=&ct \times J 
det'_{s=0}(d^\dagger d)^{-1/2}
\prod_{\sigma>0}{det'_{s=\sigma}
(d^\dagger d)^{(-1)^{\sigma+1}}}~~,
\eea
therefore recovering (\ref{Zscl}). 

\subsubsection{Regularization}

Expression (\ref{Zscl}) is of course ill defined. This is cured in standard 
manner by using zeta-function regularization. Given a 
positive elliptic operator $O$ acting on sections of the bundle ${\cal V}$, 
recall that its zeta function is defined 
through the expansion:
\be
\label{zeta}
\zeta_O(z)=\sum_{\lambda}{\frac{n_\lambda}{\lambda^z}}~~,
\ee
where $\lambda$ are the distinct eigenvalues of $O$ and $n_\lambda$ 
are their multiplicities. The series (\ref{zeta}) converges for large enough 
$Re z$, and admits a continuation to a meromorphic function defined in 
the complex plane, which is regular at the origin. If 
$\zeta'_O(z):=\frac{d\zeta_O(z)}{dz}$ denotes its derivative, then one 
defines the regularized determinant through:
\be
det^{reg}(O):=e^{-\zeta'_O(0)}~~.
\ee
Upon regularizing in this manner and using (\ref{Jf}), 
the result (\ref{Zscl}) becomes:
\be
\label{Zscl_reg}
Z_{scl}= CT(L,A_0)^{-1}\prod_{\sigma\ge 0}
[det(f_\sigma^\dagger f_\sigma)]^{(-1)^{\sigma+1}}~~,
\ee
where $C$ is a complex constant and 
we defined the {\em graded Ray-Singer torsion} by:
\be
\label{RS}
T(L,A):=
det^{',reg}_{s=0}(cd)^{1/2}\prod_{\sigma>0}{det^{',reg}_{s=\sigma}
(d^\dagger d)^{(-1)^{\sigma}}}~~.
\ee
Equations (\ref{Zscl_reg}) and (\ref{RS}) 
can be recognized as the expressions derived in 
\cite{rs}, which were obtained in that paper by using the 
method of resolvents developed in \cite{Schwarz_resolvent, Schwarz_resolvent2, 
Adams_Sen}. As expected from the general remarks of \cite{Blau_Thompson,GK}, 
this method indirectly implements the effect of ghosts and antighosts, which 
is apparent from our computations above. As explained in 
\cite{rs}, the quantity (\ref{RS}) can be used to define a `graded Ray-Singer 
norm', which is a topological invariant--a result which generalizes the 
well-known construction of \cite{Ray, RS1, RS_symp}.

\section{The effective potential}

Considering the path integral (\ref{PI}), we 
define the {\em extended potential} through:
\be
\label{Wdef}
e^{-i\lambda W_e({\hat \phi}^H)}=\frac{Z({\hat \phi}^H)}{Z(0)}=
\frac{
\int{{\cal D}[{\hat \phi}^M]
e^{-i\lambda S_e({\hat \phi}^M\oplus {\hat \phi}^H)}}}
{\int{{\cal D}[{\hat \phi}^M]
e^{-i\lambda S_e({\hat \phi}^M)}}}~~.
\ee
While this definition makes sense for arbitrary ${\hat \phi}^H$, 
the quantity $W_e$ 
has a direct physical interpretation only if 
we restrict to physical field shifts, 
${\hat \phi}^H={\hat \phi}_0^H=\phi^H$. Therefore, 
we define the {\em physical potential} $W$ by:
\be
W(\phi^H)=ev_G(W_e(\phi^H\otimes 1_G))~~{\rm~for~}\phi^H\in K^1~~.
\ee
As we shall see below, this is precisely the potential whose 
tree-level approximation was discussed in a slightly more naive language in 
\cite{gauge}. Note that (\ref{Wdef}) provides a non-perturbative 
definition of the potential, and in particular gives a prescription for its 
perturbative expansion to all loop orders. 

\subsection{Perturbative expansion} 

The perturbative expansion of $W_e$ is obtained as follows. 
First, we note that:
\bea
\label{Se_exp}
&&
S_e({\hat \phi}^M\oplus {\hat \phi}^H)=S_e({\hat \phi}^M)+
S_{e,I}({\hat \phi}^M, {\hat \phi}^H)~~,{\rm ~~with:}\\
&&S_{e,I}({\hat \phi}^M, {\hat \phi}^H)=
\frac{1}{3}\langle {\hat \phi}^H, {\hat \phi}^H*{\hat \phi}^H\rangle_e+
\langle {\hat \phi}^H, {\hat \phi}^M*{\hat \phi}^M\rangle_e+
\langle {\hat \phi}^M, {\hat \phi}^H*{\hat \phi}^H\rangle_e~~,\nn
\eea
where we used the fact that ${\hat \phi}^H$ brings no contribution to the 
kinetic term:
\be
\langle {\hat \phi}^H, d_e{\hat \phi}\rangle_e
=\langle {\hat \phi}, d_e{\hat \phi}^H\rangle_e=0~~ 
\ee
and where $S_e({\hat \phi}^M)=\frac{1}{2}\langle {\hat \phi}^M, d_e{\hat \phi}^M\rangle_e+\frac{1}{3}\langle {\hat \phi}^M, 
{\hat \phi}^M*{\hat \phi}^M\rangle_e$. Substitution of (\ref{Se_exp}) 
in (\ref{Wdef}) gives:
\be
\label{We}
e^{-i\lambda W_e({\hat \phi}^H)}=
\frac{
\int{{\cal D}[{\hat \phi}^M]
e^{-i\lambda \left[S_e({\hat \phi}^M)+S_{e,I}({\hat \phi^M}, {\hat \phi}^H)
\right]}}}
{\int{{\cal D}[{\hat \phi}^M]
e^{-i\lambda S_e({\hat \phi}^M)}}}~~.
\ee
This leads to  a perturbative series for $W_e$ upon expanding 
the exponential terms in $S_{e,I}({\hat \phi}^M,{\hat \phi}^H)$. 
Since only 
${\hat \phi}^M$ has a kinetic term, and since 
the path integral in the numerator 
is performed over this component only, this leads to Feynman integrals
in which ${\hat \phi}^H$ are treated as (amputated) external insertions,
and only ${\hat \phi}^M$ propagate. Feynman diagrams are built 
out of the vertices and propagator depicted in figure 3. We note that this 
description automatically takes ghosts and antighosts into account. 
In our formalism, their contributions are described by the non-physical 
components ${\hat \phi}^M_\sigma$ ($\sigma\neq 0$) of ${\hat \phi}^M$.

\begin{figure}[hbtp]
\begin{center}
\scalebox{0.4}{\input{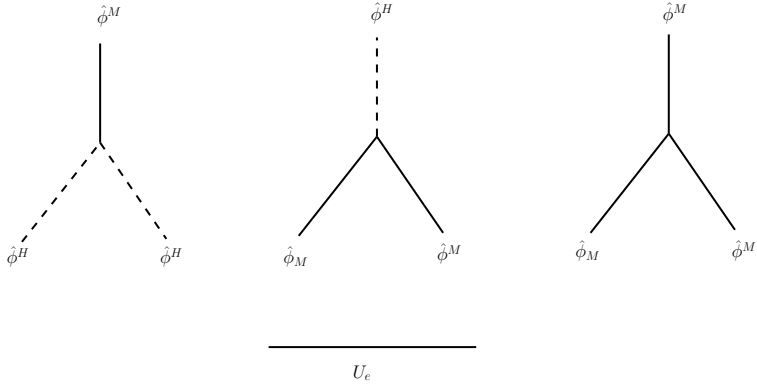}}
\end{center}
\caption{The vertices and propagator involved in the expansion of $W_e$.}
\end{figure}

\subsection{The propagator of nonzero-modes}

The propagator $U_e$ of ${\hat \phi}^M$ can be computed in 
standard manner. Proceeding as usual, we consider the free generating
functional:
\bea
\label{Zfree}
Z_{free}[{\hat L}]
&=&J\int{{\cal D}[{\hat \phi}^M]e^{-i\lambda Re
\left[\langle {\hat \phi}^M,d_e{\hat \phi}^M\rangle_e+
\langle {\hat L},{\hat \phi}^M\rangle_e\right]}}\\
&=&J\int
{{\cal D}[{\hat \phi}^M]e^{-i\lambda Re
\left[\langle{\hat \phi}^M,d_e{\hat \phi}^M\rangle_e+
\langle\pi_{d_e}{\hat L},{\hat \phi}^M\rangle_e\right]}}~~,~~~~~~~~\nn
\eea
with the external current
\footnote{Note that we do not restrict the current ${\hat L}$ to belong to 
$im d_e^\dagger$; this 
allows us to describe propagator insertions at vertices induced by 
the last two terms of $S_{e,I}$ in eq. (\ref{Se_exp}).} ${\hat L}\in {\cal H}_e^0$.
To arrive at the second form, we used the last of
properties (\ref{dder}) and the fact that ${\hat \phi}^M$ belongs to 
$im d_e^\dagger$ (together, these imply that the component of 
${\hat L}$ along $kerd_e^\dagger =K_e\oplus im d_e^\dagger $ 
does not contribute to the external coupling term).

The dependence of ${\hat L}$ can be extracted by shifting 
${\hat \phi}^M\rightarrow
{\hat \phi}^M-\frac{1}{2}d_e^{-1}\pi_{d_e}{\hat L}$, 
where $d_e^{-1}$ is the inverse 
of the  operator $d_e:im d_e^{\dagger}\rightarrow im d_e$. This gives:
\be
\langle{\hat \phi}^M,d_e{\hat \phi}^M\rangle_e+
\langle\pi_{d_e}{\hat L},{\hat \phi}^M\rangle_e\rightarrow
\langle {\hat \phi}^M,d_e{\hat \phi}^M\rangle_e-
\frac{1}{4}\langle\pi_{d_e}{\hat L},d_e^{-1}\pi_{d_e}{\hat L}\rangle_e+
\frac{1}{2}{T}~~,
\ee
where: 
\be
T=\langle {\hat \phi}^M, \pi_{d_e}{\hat L}\rangle_e-
\langle d_e^{-1}\pi_{d_e}{\hat L},  d_e{\hat \phi}^M\rangle_e=0~~,
\ee
since $\langle d_e^{-1}\pi_{d_e}{\hat L},  d_e{\hat \phi}^M\rangle_e=
\langle d_ed_e^{-1}\pi_{d_e}{\hat L},  {\hat \phi}^M\rangle_e=
\langle \pi_{d_e}{\hat L},  {\hat \phi}^M\rangle_e=
\langle {\hat \phi}^M, \pi_{d_e}{\hat L}\rangle_e$.
Thus:
\be
\label{Zf}
Z_{free}[{\hat L}]=Z_{scl}e^{\frac{i\lambda}{4} 
Re \langle \pi_{d_e}{\hat L},d_e^{-1}\pi_{d_e}{\hat L}\rangle_e}~~,
\ee
where:
\be
\label{Zfree0}
Z_{scl}=J\int{{\cal D}[{\hat \phi}^M]e^{-i\lambda Re
\langle {\hat \phi}^M,d_e{\hat \phi}^M\rangle_e}}~~.
\ee
Consider the {\em extended propagator}: 
\be
U_e:=d_e^{-1}\pi_{d_e}=\frac{1}{\Delta_e}d_e^\dagger~~.
\ee 
Then: 
$\langle \pi_{d_e}{\hat L},d_e^{-1}\pi_{d_e}
{\hat L}\rangle_e=\langle {\hat L}, U_e
{\hat L}\rangle_e$
by the last property in (\ref{dder}), since 
$im U_e=im d^\dagger$, and since $d_e^\dagger 
{\hat L}=d_e^\dagger \pi_{d_e}{\hat L}$. This allows us to re-write 
(\ref{Zf}) as:
\be
Z_{free}[{\hat L}]=Z_{scl}e^{\frac{i\lambda}{4} 
Re \langle {\hat L},U_e{\hat L}\rangle_e}~~.
\ee
Vertex insertions in the 
perturbative expansion of $W_e$ 
are now obtained by cubic functional differentiation of
$Z_{free}[{\hat L}]$ with respect to ${\hat L}$. This
also brings down insertions
of the extended propagator $U_e$.

Since $d_e=d\otimes id_G$ and $\Delta_e=\Delta\otimes id_G$, 
$U_e$ is related to the propagator $U$ of Subsection 2.2.2 by:
\be
U_e=U\otimes id_G~~.
\ee
The operators $U_e$ and $U$ have ghost degree $s=1$, and decompose as:
\be
U_e=\oplus_\sigma{U_e^\sigma}~~,~~U=\oplus_\sigma{U_\sigma}~~,
\ee
where $U_\sigma$ are linear operators from ${\cal H}(\sigma)$ to 
${\cal H}(\sigma+1)$, while $U_e^\sigma=U_\sigma \otimes id_G$ 
are operators from ${\cal H}_e(\sigma)$ to 
${\cal H}_e(\sigma +1)$. The operators $U_\sigma$ play the role of 
propagators for the various components of ${\hat \phi}$, so that 
$U$ contains both the physical propagator and the propagators of ghosts 
and antighosts.  
In this formulation, $U$ does 
not conserve the ghost number $s$ 
(since the kinetic operators $d_e$ and $d$ do
not). As in the usual Chern-Simons case,
conservation of ghost number can be achieved in the equivalent formalism 
which describes the path integral directly in terms of ghosts and antighosts 
(related to the present description 
by the change of variables explained in Subsection 4.4.)

\subsection{Feynman rules}

The Feynman rules can now be constructed in standard manner. 
In our formulation, they are virtually identical 
with the Feynman rules of usual Chern-Simons theory. 
As in that case, the vertices are only cyclically symmetric. 
This implies that the correspondence between 
Feynman integrals and graphs depends on an orientation of the plane 
(in which the graph sits). 
This is not surprising since our theories 
can be interpreted as string field theories  of oriented topological
strings \cite{sc}.
The orientation is fixed once we decide what correlation 
function corresponds to a 
graph with one vertex and three external legs and, 
of course, the same orientation
has to be used at all vertices of more complicated graphs. 

\subsection{Tree-level products}

The perturbative expansion of $W_e$ can be arranged 
according to the number of loops. In this subsection, we are interested in the 
tree-level approximation $W_e^{tree}$, which 
comes from graphs without internal circuits. An example of such a graph is 
shown in figure 4.
 
\begin{figure}[hbtp]
\begin{center}
\scalebox{0.3}{\input{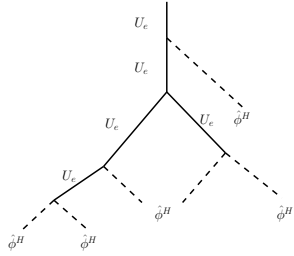}}
\end{center}
\caption{Example of a tree-level graph contributing to the perturbative 
expansion of 
$W_e$}
\end{figure}

While perturbative computation of the effective potential will generally 
require calculation of 
scattering amplitudes, the tree-level approximation $W^{tree}_e$  
can also be obtained by using 
the procedure discussed  in Appendix B of \cite{gauge}. 
The only difference between our situation and the treatment given there is 
that we are considering the extended field ${\hat \phi}$, rather than the 
classical field $\phi$.

Before we give the result, let us outline the construction. 
We begin by computing the functional  derivative 
of the extended potential $W_e$ using definition (\ref{Wdef}).
The result is that this is given by the expectation value
of the derivative of $S_e({\hat \phi}^M\oplus {\hat \phi}^H)$ 
with respect to ${\hat\phi}^H$. At this point 
we encounter the first similarity with \cite{gauge}. Using the
properties of the extended product and extended bilinear form it is 
easy to see 
that  the derivative of the extended action with respect to ${\hat\phi}^H$
is obtained from the derivative of the unextended action with 
respect to ${\phi}^H$ by replacing the unextended fields with extended 
fields and the product $\bullet$ with $*$.

Since we are interested in the tree level potential, we can use the 
saddle point approximation 
and find that we need to evaluate the derivative of the extended action on a
solution of the extended equation of motion 
$d{\hat \phi}^M+({\hat \phi}^M+{\hat \phi}^H)*({\hat \phi}^M+{\hat \phi}^H)=0$
(the critical point 
equations for $S_e({\hat \phi}^M\oplus {\hat \phi}^H)$ with respect to 
${\hat \phi}^M$). 
Since the latter is obtained from the unextended 
equation of motion by the replacement described above, so is the 
corresponding solution. 
Hence the tree level products of the extended theory are 
obtained from those of the unextended one by similar replacements.
Accordingly, we can use Appendix B of \cite{gauge} to conclude that 
the extended tree-level potential can be written in the form:
\be
\label{extW}
W_e^{tree}=-
\sum_{n\geq 2}{\frac{(-1)^{n(n+1)/2}}{n+1}
\lambda^{n-2}\langle {\hat \phi}^H, r_n^e(({\hat \phi}^H)^{\otimes n})
\rangle_e}~~,
\ee
where: 
\be
r_n^e(({\hat\phi}^H)^{\otimes n})=P_e\lambda_n^e(({\hat\phi}^H)^{\otimes n})
\ee
and $P_e$ is the projector on $K_e$. Exactly as in \cite{gauge}, 
the products $\lambda_n^e$ satisfy the
recurrence relations:
\bea
\label{recl}
\lambda^e_2(({\hat\phi}^H)^{\otimes 2})&=&{\hat\phi}^K * {\hat \phi}^K~~,\\
{\lambda}^e_n (({\hat\phi}^H)^{\otimes n}) &=& 
(-1)^{n-1} {\hat\phi}^H * U_e{\lambda}^e_{n-1}(({\hat\phi}^H)^{\otimes (n-1)}) + 
 (-1)^{n-1}
U_e{\lambda}^e_{n-1}(({\hat\phi}^H)^{\otimes (n-1)}) * {\hat\phi}^H \nn\\
 &-&
\sum_{\stackrel{l + m = n}{l,m\geq 2}} (-1)^{m l}U_e{\lambda}^e_l(({\hat\phi}^H)^{\otimes l}) * 
U_e{\lambda}^e_m (({\hat\phi}^H)^{\otimes m}) ~~{\rm for}~~
n\geq 3 ~~.\nn
\eea

The extended potential constructed here depend on 
physical fields as well as ghosts and antighosts.
To isolate the potential for physical fields we particularize to
harmonic shifts of the 
form ${\hat\phi}^K=\phi^K\otimes 1_G$, where $\phi_K\in {\cal H}^1$. 
As explained in Subsection 3.1., 
the products $*$ and $\bullet$ agree for such fields --- in fact, we 
have $(u\otimes 1_G)*(v\otimes 1_G)=(u\bullet v)\otimes 1_G$, since the unit 
$1_G$ of the Grassmann algebra is even. 
Combining this with 
the properties $U_e=U\otimes id_G$, $P_e=P\otimes id_G$, we obtain:
\be
\lambda^e_n((\phi^H\otimes 1_G)^{\otimes n})=
\lambda_n((\phi^H)^{\otimes n})\otimes 1_G~~,~~
r^e_n((\phi^H\otimes 1_G)^{\otimes n})=r_n((\phi^H)^{\otimes n})\otimes 1_G~~.
\ee
These relations follow by a simple induction argument, upon using the 
fact that $\phi^H$ has even Grassmannality and the definition (\ref{star}) 
of the extended product. 

This allows us to recover the tree-level potential $W^{tree}$ 
constructed in \cite{gauge}:
\be
W^{tree}_e(\phi^H\otimes 1_G)=W^{tree}(\phi^H)\otimes 1_G~~,
\ee
where we used definition (\ref{extended_form}) of the extended bilinear form. 

\section{Conclusions}

We discussed gauge-fixing of graded Chern-Simons field theories. Upon 
choosing a `universal' delta-function gauge fermion, we showed that the 
resulting gauge-fixed action leads to a formalism which is 
a graded version of the description given in \cite{AS1}. This leads to 
a simple expression for the propagator of nonzero-modes, and allows for 
a systematic treatment of the semiclassical approximation. In particular, 
we re-derived the semiclassical partition function of \cite{rs} within
the BV formalism, obtaining complete agreement with the results predicted by 
the method of resolvents. Finally, we discussed the perturbative expansion
of the effective potential for zero modes, and provided the BV justification of
the construction used in \cite{gauge}.

The current work (and, more generally, our interest in graded Chern-Simons 
theories) is motivated by their role in a 
string-field theoretic description of 
topological D-branes in Calabi-Yau compactifications. In the present context,
this arises when the 3-manifold $L$ is a special Lagrangian 3-cycle of a 
Calabi-Yau threefold. Regarding this connection, we mention that there exists a
B -model version of our theories \cite{Diaconescu}, which is in many ways 
extremely similar; we chose to focus on the A-model mostly due to its 
nontrivial topological implications. 

As discussed in \cite{gauge}, the graded Chern-Simons description allows 
for a natural definition of the moduli space of graded D-branes,  
which does not seem to admit a simple treatment in other approaches. From 
that perspective, it is interesting to consider 
the role of our potential and semiclassical approximation at the level 
of the associated triangulated categories \cite{sc, Diaconescu}. 
This and related issues are currently under investigation.

\acknowledgments{
We are indebted to M.~Rocek
for support and interest in our work.  C.I. L. thanks Rutgers University 
(where part of this paper was completed) 
for hospitality and providing excellent conditions.
The present work was supported by the Research Foundation under NSF grants 
PHY-9722101, NSFPHY00-98395 (6T) and by the DOE grant 91ER40618 (3N).
}

\appendix

\section{A weighted gauge} 

One can choose an alternate gauge fixing fermion for our systems and show that 
that  the extra ghosts decouple in the gauge-fixed action; 
the corresponding family of weighted gauges 
recovers the Landau gauge of Section 4 in an appropriate limit. 
The present appendix gives a general discussion of this gauge as well as an 
independent treatment for an example, showing how the weighted 
gauge-fixing procedure can be recovered by more elementary means in a 
particular case.
For simplicity, we start once again with the acyclic case.

\subsection{The acyclic case}
 
For an acyclic background, the weighted gauge-fixing fermion has the form:
{\footnotesize \bea
\label{fermion_weighted}
\Psi_N&{}&=Re\int_{L} str_e\sum_{k\geq 0}\left[ \sum_{\sigma\geq 2k} 
{{\overline \phi}^N_{\sigma+1}[k]*d_ec_e\phi^N_\sigma[k]}+\sum_{\sigma\geq
2k+1} {\phi^N_{\sigma+1}[k+1]* d_ec_e{\overline \phi}^N_\sigma[k]} \right. 
\nonumber\\
&{}&~~~~~~~~~~~~~~~~ + \Xi  
\sum_{\sigma\ge 2 k}{\overline\phi}^N_{\sigma+2}[k] * c_e\pi^N_{\sigma+2}[k+1]+
\Xi
\sum_{\sigma\ge 2 k+1}\phi^N_{\sigma+1}[k+1]*c_e{\overline \pi}^N_{\sigma+1}[k]
\nonumber\\
&{}&~~~~~~~~~~~~~~~~\left.
+\Xi~{\overline\phi^N}_{2k+1}[k]*c_e{\overline \pi}^N_{2k+1}[k]
\right]~~,
\eea}\noindent 
where the positive constant $\Xi$ is the weight. 
The antifields are now eliminated through the equations:
{\footnotesize \bea
\label{eq_elim1}
\phi^{N*}_\sigma[k]&=& c_ed_e{\overline\phi}^N_{\sigma+1}[k]-
d_ec_e{\overline\phi}^N_{\sigma-1}[k-1]-\Xi~
c_e{\overline \pi}^N_\sigma[k-1]~~\label{eq:weightedantif}\\ 
{\overline\phi^N}^*_\sigma[k]&=& c_ed_e\phi^N_{\sigma+1}[k+1]- 
d_ec_e\phi^N_{\sigma-1}[k]-\Xi ~
c_e\pi^N_\sigma[k+1]-\Xi~\delta_{\sigma,\,2k+1}c_e{\overline \pi}^N_{2k+1}[k]
\nonumber~~.
\eea}
For $k=0$, the first relation gives:
\be
\phi^{N*}_\sigma=c_ed_e{\overline\phi}^N_{\sigma+1}~~,
\ee
which reproduces the gauge-fixing condition of Section 4.
Note that the auxiliary fields $\pi^N$ and ${\overline \pi}^N$
will not appear in $S_{e,gf}$ once we replace the antifields by the equations 
above. Indeed, the only antifields appearing in $S_e$ are 
$\phi^{N*}_\sigma[0]$ and they do not depend on $\pi^N$ and ${\overline \pi}^N$.

Substituting (\ref{eq:weightedantif}) in the auxiliary action and 
using the equations of motion to eliminate the auxiliary fields $\pi^N$ and ${\overline \pi}^N$, we find: 
{\footnotesize \bea
&&\!\!\!\!S_{aux,gf}={1\ov 4\Xi}Re\sum_{k\ge 0}
\langle c_ed_e\phi^N_{2(k+1)}[k+1]-d_ec_e\phi^N_{2 k}[k],\,
d_e\phi^N_{2(k+1)}[k+1]-c_ed_ec_e\phi^N_{2 k}[k]\rangle_e\\
&&\!\!\!\!+{1\ov 2\Xi}Re \sum_{k\ge 1}\sum_{\sigma\ge 2k}(-1)^{\sigma+1}
\langle c_ed_e\phi^N_{\sigma+1}[k]-d_ec_e\phi^N_{\sigma-1}[k-1],\,
(-1)^{\sigma+1}d_e{\overline\phi}^N_{\sigma+1}[k]+c_ed_ec_e
{\overline\phi}^N_{\sigma-1}[k-1]\rangle_e ~~.\nonumber
\eea}
Integration by parts gives the equivalent form:
{\footnotesize \bea
S_{aux,gf}=&&-{1\over 4 \Xi} 
Re\langle \phi^N_{0}[0],\,c_ed_ec_ed_ec_e\phi^N_{0}[0]\rangle_e
-{1\over 4 \Xi} Re\sum_{k\ge 1}
\langle \phi^N_{2k}[k],(dcd+\,c_ed_ec_ed_ec_e)\phi^N_{2k}[k]\rangle_e\nn\\
&&
+{1\over 2 \Xi} Re\sum_{k\ge 1}\sum_{\sigma\ge 2k+1}
\langle \phi^N_{\sigma}[k] ,\,\left[
d_ec_ed_e+
(-1)^\sigma c_ed_ec_ed_ec_e \right]
{\overline\phi^N}_{\sigma}[k]  \rangle_e
\nonumber\\
&&
+{1\over 2 \Xi} Re\sum_{\sigma\ge 2}(-1)^\sigma
\langle\phi^N_{\sigma}[0],\,c_ed_ec_ed_ec_e
{\overline\phi}^N_{\sigma}[0]\rangle_e~~,
\eea}\noindent 
where we used $c_e^2{\hat u}=(-1)^{g({\hat u})}{\hat u}$. 
This expression shows that
the extraghosts are free and thus can be integrated out. 
This produces an appropriate prefactor in the 
path integral measure and allows us to replace $S_{aux}$ by its first term:
{\footnotesize \be
S_{aux,fd}\equiv -{1\over 4 \Xi} 
Re\langle \phi^N_{0},\,c_ed_ec_ed_ec_e\phi^N_{0}\rangle_e=
~~-{1\over 4 \Xi} 
Re\langle \phi^N_{0},\,c_ed_ed_e^\dagger\phi^N_{0}\rangle_e=
{1\over 4 \Xi} 
Re h_e(d_e^\dagger\phi^N_{0},\,d_e^\dagger\phi^N_{0})~~.
\ee}
Hence the result of gauge-fixing can be described through
the action:
\be
S_{gf}^\Xi=S_e(\phi^{N*}_\sigma=c_ed_e{\overline\phi}^N_{\sigma+1}, \phi^N_\sigma) 
+{1\over 4 \Xi} 
Re h_e( d_e^\dagger \phi^N_{0},\,d_e^\dagger\phi^N_{0})~~.
\ee

The gauge of Section 4 is now obtained in the Landau-type
limit\footnote{Since our systems are defined on the {\em compact} 
manifold $L$, 
we do not have to worry about infrared singularities when taking 
this limit. Compacity 
of $L$ acts as an automatic infrared regulator, which we never have to 
remove. }  
$\Xi\rightarrow 0$. 
As usual, this can be implemented at the level of the action
by dropping $S_{aux}$ and keeping in mind that all propagators 
need to be transverse.
Alternately, one can take this limit at 
the level of Feynman rules (in which case transverse 
propagators will be produced automatically). This reproduces the 
Landau gauge. 

\subsection{Inclusion of zero-modes} 

In the presence of zero-modes, we supplement the gauge-fixing fermion
by adding the harmonic piece:
{\footnotesize \bea 
\Psi_K&=&
Re\int_{L}str_e\sum_{k\geq 0}\left( \sum_{\sigma\geq 2k}{ 
(c_e{\overline\mu}^K_{\sigma}[k])*A_\sigma[k]^{1/2}\phi^K_\sigma[k]}+ 
\sum_{\sigma\geq 2k+1} {(c_e\mu^K_{\sigma}[k])* A_\sigma[k] ^{1/2}
{\overline \phi}^K_\sigma[k]}\right)\nonumber\\ 
&+&\Xi~ Re\int_{L}str_e\sum_{k\geq 0} \left( 
\sum_{\sigma\ge 2 k}A_{\sigma+2}[k] ^{1/2}
{\overline\phi}^K_{\sigma+2}[k] * c_e 
\pi^K_{\sigma+2}[k+1]+\right. \\ 
&+&\left.
\sum_{\sigma\ge 2 k+1}A_{\sigma+1}[k+1]^{1/2}\phi^K_{\sigma+1}[k+1]* 
c_e{\overline\pi}^K_{\sigma+1}[k] 
+~A_{2k+1}[k]^{1/2}{\overline\phi}^K_{2k+1}[k]* 
c_e{\overline\pi}^K_{2k+1}[k]\right)\nonumber~~. 
\eea}\noindent  
where $A_\sigma[k]$ are, as in Section 4.2.2, some strictly positive operators 
on $K(\sigma)$.
This fixes the harmonic components of antifields to the values:
{\footnotesize \bea
&&\phi^K{}^*_\sigma[k]=\frac{\delta \Psi}{\delta \phi^K_\sigma[k]} 
=A_\sigma[k]^{1/2}c_e{\overline\mu}^K_\sigma[k]
+\Xi\delta(\stackrel{k\ge 1}{\scriptstyle\sigma\ge 2k})
A_\sigma[k]^{1/2}c_e{\overline\pi}^K_\sigma[k-1] \\
&&{\overline \phi}^K{}^*_\sigma[k]= 
\frac{\delta \Psi}{\delta {\overline\phi}^K_\sigma[k]} 
=A_\sigma[k]^{1/2}c_e{\mu}^K_\sigma[k]+
\Xi\delta(\stackrel{k\ge 0}{\scriptstyle\sigma\ge 2k+2})
A_\sigma[k]^{1/2}c_e{\pi}^K_\sigma[k+1]\nn\\
&&~~~~~~~~~~~~~~~~~~~~~~~~~~~~~~~~~~~~~~~~~~~+
\Xi\delta_{\sigma,2k+1}\delta{\scriptstyle k\ge0}
A_{2k+1}[k]^{1/2}c_e{\overline\pi}^K_{2k+1}[k]\nn\\
&&\mu^K{}^*_\sigma[k]=\frac{\delta \Psi}{\delta \mu^K_\sigma[k]} 
=c_eA_\sigma[k]^{1/2}{\overline\phi}_\sigma^K[k] ~~;~~~~~
{\overline\mu}^K{}^*_\sigma[k]=\frac{\delta \Psi}{\delta  
{\overline\mu}^K_\sigma[k]}=c_eA_\sigma[k]^{1/2}\phi_\sigma^K[k]~~. \nn
\eea}\noindent 
Substituting  them in the auxiliary action gives:
{\footnotesize \bea 
S_{aux}&=& 
Re\int_{L}str_e\left(\sum_{k\geq 0}\sum_{\sigma\geq 2k}
\lambda^K_\sigma[k]* 
A_\sigma[k]c_e{\phi}_\sigma^K[k]\right.  \\
&{}&\left.~~~~~~~~~~~~~~~~+ 
\sum_{k\geq 0}\sum_{\sigma\geq 2k+1} 
{\overline \lambda}^K_\sigma[k]*
A_\sigma[k]c_e{\overline\phi}_\sigma^K[k]
\right) \nonumber
\\ 
&+&Re\int_{L}str_e\left(\sum_{k\geq 1}\sum_{\sigma\geq 2k}{ 
\pi^K_\sigma[k]*A_\sigma[k]c_e({\overline\mu} 
_\sigma^K[k]+\Xi{\overline\pi}^K_\sigma[k-1])}\right.\nonumber\\
&{}&\left.~~~~~~~~~~~~~~~~+\sum_{k\geq 0}\sum_{\sigma\geq 2k+2}{ 
{\overline \pi}^K_\sigma[k]*A_\sigma[k] 
c_e(\mu^K_\sigma[k]+\Xi{\pi}^K_\sigma[k+1])}\right.\nn\\
&{}&\left.~~~~~~~~~~~~~~~~+\sum_{k\geq 0}
{\overline \pi}^K_{2k+1}[k]* 
A_\sigma[k]c_e(\mu^K_{2k+1}[k]+\Xi{\overline\pi}^K_{2k+1}[k])
\right) \nonumber~~.
\eea} 

The integral over $\lambda$ and ${\overline\lambda}$ gives the same 
result as before. For the remaining variables, we 
first integrate over $\pi[k]$ and find
{\footnotesize \be
\prod_{\stackrel{k\ge 1}{\sigma\ge 2k}}\delta\left(
A_\sigma[k]c_e({\overline\mu} 
_\sigma^K[k]+\Xi{\overline\pi}^K_\sigma[k-1])+\Xi c_eA_\sigma[k]
{\overline\pi}_\sigma[k-1]\right)~~.
\ee}\noindent
Now integration over ${\overline\mu}$ produces the first factor 
in (\ref{rest}). The remaining integrals produce the second factor
in that equation. 

\subsection{An example}

Let us consider an example  related to 
the `D-brane pairs of unit relative grade', 
which were discussed from various points of view in 
\cite{bv, gauge, rs}. 
For the reader's convenience, 
we give a direct construction of the gauge-fixing 
fermion discussed on general grounds above. 
For simplicity, we shall assume an acyclic background. 

The setup consists of two graded flat bundles (of equal ranks) 
$E_a$ and $E_b$ on $L$, 
whose grades differ by one: $grade(b)=grade(a)+1$.
A generic field configuration degree $|u|=n$ is represented by the matrix
\be
u=\left[\begin{array}{cc} u^{(n)}_{aa}~~~&~~~u^{(n+1)}_{ba}\\
    u^{(n-1)}_{ab}&u^{(n)}_{bb}\end{array}\right]~~,
\label{eq:morphisms}
\ee 
with: 
\bea u^{(1)}_{aa}\in
Hom^1(a,a)=\Omega^1(End(E_a))&,&u^{(1)}_{bb}\in Hom^1(b,b)=
\Omega^1(End(E_b))\nn\\
u^{(1-n)}_{ab}\in Hom^1(a,b)=\Omega^{1-n}(Hom(E_a,E_b)) &,&
u^{(1+n)}_{ba}\in Hom^1(b,a)=\Omega^{1+n}(Hom(E_b,E_a))\nn~,~~~~~
\eea 
where the superscripts in round brackets indicate form rank. 
This is the notation 
used in \cite{bv}. The 
physical fields $\phi_0[0]$ and first and second generation 
ghosts $\phi_1[0]$  and  $\phi_2[0]$ have the form:
\be
\phi_0[0]=\pmatrix{\phi^{(1)}&\phi^{(2)}\cr
\phi^{(0)}&\phi{}'^{(1)}\cr}~~,~~
\phi_1[0]=\pmatrix{c_1^{(0)}&c_1^{(1)}\cr
0&c'{}_1^{(0)}\cr}~~,~~
\phi_2[0]=\pmatrix{0&{ c}{}_2^{(0)}\cr
0&0\cr}~~.
\ee
For each matrix entry, the upper index denotes 
form rank while the lower index is the generation number.
As mentioned above, we shall assume an acyclic background. 
As explained in \cite{gauge, bv}, this can be achieved by 
giving an expectation value to the field $\phi^{(0)}$, such 
that this vev is flat bundle isomorphism. 

To construct the quantum action we add a trivial pair for each 
classical gauge invariance and write the non-minimal action:
\be
S_{nm}=Re\,\int\, tr[\pi{}_1^{(0)} b^*{}_1^{(3)}+
\pi'{}_1^{(0)} b'{}_1^*{}^{(3)} + \pi{}_1^{(1)} b{}^*{}_2^{(2)}
+{\pi}{}_2^{(0)} { b}{}^*{}^{(2)}
+...]
\ee
where juxtaposition stands for the total boundary product $\bullet$ and 
the dots represent 
possible extra terms due to invariances of the gauge fixing fermion.
The weighted gauge-fixing fermion has the form:
\bea
&&\Psi=Re
\int\, tr[b{}_1^{(0)}d*\phi{}^{(1)}+b'{}_1^{(0)}d*\phi'{}^{(1)}
+b{}_1^{(1)}d*\phi{}^{(2)}+b{}_2^{(0)}d*c{}_1^{(1)}+ {\check c}{}_2^{(0)}
d*b{}_1^{(1)}\nn\\
&&+\Xi[b{}_1^{(0)}*\pi{}_1^{(0)}+b'{}_1^{(0)}*\pi'{}_1^{(0)}
+b{}_1^{(1)}*\pi{}_1^{(1)}+b{}_2^{(0)}*{\check \pi}{}_2^{(0)}
+ {\check c}{}_2^{(0)}*{ \pi}{}_2^{(0)}]~~.
\label{eq:compgff}
\eea
One could be tempted to 
write $b{}_2^{(0)}*{\pi}{}_2^{(0)}$ instead of the second last 
term. However, such a term is incompatible with the 
ghost number requirements for $\Psi$ 
and $S_{nm}$.
Furthermore, the gauge invariance $b{}_1^{(1)}\rightarrow 
b{}_1^{(1)}+ d\kappa$
requires the introduction of an extra ghost ${\check c}$. By counting the 
ghost numbers associated with the various fields we find that 
the last term in the previous equation has ghost number $-1$ and thus it is
indeed allowed in $\Psi$. Due to the presence of these extra fields 
we need to add a further term to the non-minimal action:
\be
S_{extra}=Re\,\int \,tr[{\check \pi}{}_2^{(0)}{\check c}{}^*{}_2^{(3)}]~~.
\ee
This satisfies the ghost number and parity constraints provided that 
${\check c}$ and ${\check \pi}$ form a trivial pair. The result is the 
full auxiliary action $S_{aux}=S_{nm}+S_{extra}$.

The auxiliary fields can be arranged in the following matrices:
\bea
\bar\pi_1[0]=\pmatrix{-\pi_1^{(0)}&-\pi_1^{(1)}\cr 0&-\pi'{}_1^{(0)}\cr}~~&,&~~
\bar \phi_1[0]= \pmatrix{-b_1^{(0)}& b_1^{(1)}\cr 0& -b'{}_1^{(0)}\cr}~~\nn\\
\pi_2[1]=~~~~\pmatrix{0& \pi{}_2^{(0)}\cr0&0\cr}~~&,&~~
\bar \phi_2[0]=\pmatrix{0& b_2^{(0)}\cr 0&0}~~\nn\\
\phi_2[1]=~~~~\pmatrix{0&\check c_2^{(0)}\cr0&0\cr}~~&,&
~~{\bar \pi}_2[0]=\pmatrix{0&{\check \pi}{}_2^{(0)}\cr 0& 0\cr}~~.
\eea
This presentation allows us to write the gauge-fixing fermion and 
auxiliary action in the form given in the previous subsections:
\bea
\Psi=&&Re\,\int \,
str\left[{\bar\phi}_1[0]*d_ec_e\phi_0[0]+{\bar\phi}_2[0]*d_ec_e\phi_1[0]
+{\phi}_2[1]*d_ec_e{\bar \phi}_1[0]\right.\nn\\
&&\left.+\Xi[{\bar\phi}_1[0]*c_e{\bar \pi}_1[0]+
{\bar\phi}_2[0]*c_e\pi_2[1] + \phi_2[1]*c_e{\bar\pi}_2[0]]\right]
\eea
and:
\bea
S_{aux}&&=S_{nm}+Re\int\,tr[{\check \pi}{}_2^{(0)}{\check c}{}^*{}^{(3)}]\\
&&=Re
\int \,tr[{\bar\pi}_1[0]*{\bar\phi}_1^*[0]+{\bar\pi}_2[0]*{\bar\phi}_2^*[0]
+\pi_2[1]*\phi_2^*[1]]~~.\nn
\eea
where
\be
{\overline\phi}_1^*[0]=\pmatrix{-b^{*(3)}_1 & 0\cr -b^{*(2)}_1 &-
{b'}^{*(3)}_1 }
~~~~~~~~
{\overline\phi}_2^*[0]=\pmatrix{0&0\cr b^{*(3)}_2&0}
~~~~~~~~
{\phi}_2^*[1]=\pmatrix{0&0\cr {\check c}^{*(3)}_2&0}
\ee

Following standard procedure, we solve 
for the transformed antifields and substitute the solutions into
the full action $S_{tot}=S_e+S_{aux}$. 
We first analyze the terms related to 
gauge symmetries of the two-form.
The relevant terms in $S_{tot}$ are:
\bea
S_{tot}&&=\int tr\left[\dots+\phi^{(0)} d\phi^{(2)}+\phi^*{}^{(1)}
(d c_1^{(1)}+\dots )+ c^*{}_1^{(2)} (d { c}_2^{(0)}+\dots)\right.\\
&&\left. + \pi_1^{(1)} b^*{}_1^{(2)}+{\pi}{}_2^{(0)} 
{b}^*{}_2^{(3)}+{\check \pi}_2^{(0)} {\check c}^*{}_2^{(3)}\right]+\dots~~.
\eea
Solving for the antifields from the gauge fixing fermion leads to the 
gauge-fixed action:
\bea
S_{gf}&=&Re\,\int\,tr\left[\dots+\phi^{(0)} d\phi^{(2)}
+ *  db{}_1^{(1)} (d_ec{}_1^{(1)}+\dots) - *  d{ b}{}_2^{(0)}
(d{c}_2^{(0)}+\dots)\right.\nonumber\\
&+&\pi_1^{(1)} (\Xi * \pi_1^{(1)} + d * 
\phi^{(2)}- *  d{\check c}{}_2^{(0)}) + 
{ \pi}{}_2^{(0)} (\Xi  *  {\check \pi}_2^{(0)}
+ d *  c_1^{(1)})\nonumber\\
&+&\left.{\check\pi}_2^{(0)} (\Xi *  {\pi}_2^{(0)}
+d *  b_1^{(1)})\right]+\dots~~.
\eea
After eliminating the auxiliary fields and dropping 
total derivatives, we obtain:
\bea
S_{gf}&=&Re\int\,tr\left[\dots+\phi^{(0)} d\phi^{(2)}
+ *  db{}_1^{(1)} (d c{}_1^{(1)}+\dots) - *  d{ b}{}_2^{(0)}
(d{c}{}_2^{(0)}+\dots)\right.\nonumber\\
&-&\frac{1}{4\Xi}( *  d{\check c}{}_2^{(0)}
d{\check c}{}_2^{(0)} + d * \phi^{(2)}  * 
d * \phi^{(2)} )-\left.{1\ov 2\Xi}d *  c_1^{(1)}*d *  b_1^{(1)}\right]~~.
\eea 
where we used the fact that ${ \pi}{}_2^{(0)}$ and ${\check\pi}_2^{(0)}$
are anti-commuting objects.
The last expression shows that 
the gauge-fixing conditions are actually decoupled and have  the
Lorentz form $d^\dagger \phi^{(2)}= d^\dagger c^{(1)}= d^\dagger b_{(1)}=0$.
The same condition can be derived from the 
delta function gauge fixing fermion. Then, the auxiliary fields appear 
as Lagrange multipliers 
imposing the various gauge constraints. For example,
$\pi^{(1)}$ imposes 
$*d*\phi^{(2)}= d\check c^{(0)}$.
This implies that $\check c^{(0)}$ is a harmonic
form. Since the base manifold $L$ is compact, 
one concludes that $\check c^{(0)}$ is a covariantly-constant
section on $End(E)$, and that $\phi^{(2)}$ also satisfies 
the gauge condition $d^\dagger \phi^{(2)}=0$.

Some straightforward algebra leads to the complete form of 
the gauge-fixed action:
{\footnotesize \bea S_{gf}&&
=2Re\left\{ft\int_L tr_a\left[
  {1\ov 2}\left(\p{1}d\p{1}-\p{2}d\p{0}\right)+{1\ov
    3}\left(\p{1}\p{1}\p{1}+
    \p{1}\p{2}\p{0}\right)\right]~~\right. \nn \\
&&-\int_{L}tr_b\left[{1\ov
    2}\left(\hp{1}d\hp{1}-\p{0}d\p{2}\right)+{1\ov 3}\left(
    \hp{1}\hp{1}\hp{1}+\hp{1}\p{0}\p{2}\right)\right]~~\nn\\
&&+\int_L\,tr_a\Big[~
b_1^{(0)}d*
\left(dc_1^{(0)}+[\p{1},\,c_1^{(0)}]-c_1^{(1)}\p{0}\right)
\nonumber\\
&&+\int_L\,tr_b\Big[~ b{}_1^{(1)}d*\left(
  dc_1^{(1)}+(\p{1}c_1^{(1)}+\p{2}{c'}_1^{(0)}-c_1^{(0)}\p{2}+
  c_1^{(1)}\hp{1}) \right)\nn\\
&&~~~~~~~~+
b'{}_1^{(0)}d*\left(d{c'}_1^{(0)}+[\hp{1},\,{c'}_1^{(0)}]-\p{0}c_1^{(1)}\right)
\nonumber\\
&&~~~~~~~~+
b_2^{(0)}d*
\left(c_1^{(0)}c_1^{(1)}+c_1^{(1)}{c'}_1^{(0)}+dc_2^{(0)}+
\p{1}c_2^{(0)}-  c_2^{(0)}\hp{1}\right)\Big]
\nonumber\\
&&+\int_L\,tr_a\left(-c_2^{(0)}*db{}_1^{(1)} *db_1^{(0)}+
  c_2^{(0)}*db'{}_1^{(0)}*db{}_1^{(1)}\right) \nn\\
&&- \frac{1}{4\Xi}\int_L\,
tr_a\Big[d*\phi^{(1)}*d*\phi^{(1)}+d * \phi^{(2)}  * 
d * \phi^{(2)}\nn\\
&&- \left.\frac{1}{4\Xi}\int_L\,
tr_b\Big[d*\phi'{}^{(1)}*d*\phi'{}^{(1)}+ *d{\check c}{}_2^{(0)}
d{\check c}{}_2^{(0)}+2d *  c_1^{(1)}*d *  b_1^{(1)}\Big]\right\}~~.
\eea}
This expression can be recognized as a particular case of (\ref{Sgf}):
{\footnotesize \bea
S_{gf}&&=2Re\left\{\int_L\,str\Bigg[
{1\ov 2}\phi_0[0]*d_e\phi_0[0]+{1\ov 3}\phi_0[0]*\phi_0[0]*\phi_0[0]\right.
\nn\\
&&~~~~~~~~~~~~
+c_ed_e{\bar \phi}_1[0]*(d\phi_1[0]+[\phi_0[0],\,\phi_1[0]]_*)\nn\\
&&~~~~~~~~~~~~
+c_ed_e{\bar \phi}_2[0]*(d\phi_2[0]+
[\phi_0[0],\,\phi_2[0]]_*+[\phi_1[0],\,\phi_1[0]]_*)
\nn\\
&&~~~~~~~~~~~~
+\phi_2[0]*[c_ed_e{\bar \phi}_1[0],\,c_ed_e{\bar \phi}_1[0]]_*     \\
&&~~~\left.
-{1\ov 4\Xi}\left(\phi_2[1]*d_ec_ed_e\phi_2[1]+\phi_0[0]*c_ed_ec_ed_ec_e
\phi_0[0]+
2\phi_1[0]*c_ed_ec_ed_ec_e{\bar\phi}_1[0]\right)
\Bigg]\right\}\nn~~.
\eea}
The first line 
is the classical action, the next two lines come 
from terms linear in antifields, the fourth line is produced by 
terms quadratic in antifields, while the remaining contributions come from 
$S_{aux}$.
We also used the fact that $d^\dagger_e\phi_2[1]=0$ and 
$d^\dagger_e{\overline\phi}_2[0]=0$ which follow
from the explicit form of $\phi_2[1]$ and ${\overline\phi}_2[0]$.


\begin{thebibliography}{100}

\bibitem{Douglas_Kontsevich}{M.~R.~Douglas, {\em D-branes, Categories
      and N=1 Supersymmetry}, hep-th/0011017.}  

\bibitem{Aspinwall}{P.~S.~Aspinwall,
    A.~Lawrence, {\em Derived Categories and Zero-Brane Stability},
    hep-th/0104147.}

\bibitem{Douglas_Aspinwall}{P.~S.~Aspinwall, M.~R.~Douglas, 
{\em D-Brane Stability and Monodromy}, hep-th/0110071.}

\bibitem{com1}{
    C.~I.~Lazaroiu, {\em Generalized complexes and string field
    theory}, JHEP 06 (2001) 052.} 

\bibitem{com3}{C.~I.~Lazaroiu, {\em
    Unitarity, D-brane dynamics and D-brane categories},
    hep-th/0102183. } 

\bibitem{sc}{C.~I.~Lazaroiu, {\em Graded
    Lagrangians, exotic topological D-branes and enhanced triangulated
    categories}, JHEP 0106 (2001) 064.}

\bibitem{Diaconescu}{D.~E.~Diaconescu, {\em
    Enhanced D-Brane Categories from String Field Theory},
    hep-th/0104200.}  

\bibitem{gauge}{ C. I. Lazaroiu, R. Roiban, {\em 
Holomorphic potentials for graded D-branes},  hep-th/0110288.} 
\bibitem{rs}{C. I. Lazaroiu, {\em An analytic torsion for graded D-branes},
hep-th/0111239.}

\bibitem{Witten_CS}{ E.~Witten,{\em
    Chern-Simons gauge theory as a string theory}, The Floer memorial
    volume, 637--678, Progr. Math., 133, Birkhauser, Basel, 1995,
    hep-th/9207094.}  

\bibitem{Witten_nlsm}{ E.~Witten, {\em Topological sigma models},
    Commun. Math. Phys.  {\bf 118} (1988),411.}
\bibitem{Witten_mirror}{E.~Witten, {\em Mirror manifolds and
      topological field theory}, Essays on mirror manifolds, 120--158,
    Internat. Press, Hong Kong, 1992, hep-th/9112056.}

\bibitem{Witten_antibracket}{E.~Witten, {\em A note on the antibracket
      formalism}, Mod. Phys. Lett. {\bf A5} (1990) 487.}
\bibitem{Henneaux_geom}{M.~Henneaux, {\em Geometric Interpretation of
      the Quantum Master Equation in the BRST--anti-BRST Formalism},
    Phys. Lett. {\bf B282} (1992) 372, hep-th/9205018.}
\bibitem{Khudaverdian}{O.~M.~Khudaverdian, A.~P.~Nersessian, {\em On
      the Geometry of the Batalin-Vilkovisky Formalism}, Mod. Phys.
    Lett. {\bf A8} (1993) 2377-2386, hep-th/9303136.}
\bibitem{Schwarz_geom}{A.~Schwarz, {\em Geometry of Batalin-Vilkovisky
      quantization}, Commun.Math.Phys. {\bf 155} (1993) 249-260.}
\bibitem{Schwarz_semiclassical}{A.~Schwarz, {\em Semiclassical
      approximation in Batalin-Vilkovisky formalism},
    Commun.Math.Phys. {\bf 158} (1993) 373-396.}
\bibitem{Schwarz_symms}{A.~Schwarz, {\em Symmetry transformations in
      Batalin-Vilkovisky formalism}, Lett.Math.Phys. {\bf 31} (1994)
    299-302.}  \bibitem{Schwarz_superanalogues}{A.~Schwarz, {\em
    Superanalogs of symplectic and contact geometry and their
    applications to quantum field theory}, hep-th/9406120.}
 \bibitem{Kontsevich_Schwarz}{ M.
    Alexandrov, M. Kontsevich, A. Schwarz, O. Zaboronsky, {\em The
    Geometry of the Master Equation and Topological Quantum Field
    Theory}, Int.J.Mod.Phys. {\bf A12} (1997) 1405-1430,
    hep-th/9502010.}  
\bibitem{AS1}{S.~Axelrod, I.~M.~Singer, {\em Chern-Simons perturbation
      theory}, Proceedings of the XXth International Conference on
    Differential Geometric Methods in Theoretical Physics, Vol. 1, 2
    (New York, 1991), 3--45, World Sci. Publishing, River Edge, NJ,
    1992, hep-th/9110056.} 

\bibitem{Gomis}{J.~Gomis,
    J.~Paris, S.~Samuel, {\em Antibracket, Antifields and Gauge-Theory
    Quantization}, Phys.Rept. {\bf 259} (1995) 1-145.}

\bibitem{Nash_Connor}{C.~Nash, D.~O' Connor, 
{\em BRST Quantization and the Product Formula for the Ray-Singer Torsion},
Int.J.Mod.Phys. {\bf A10} (1995) 1779-1806, hep-th/9310038.}

\bibitem{Schwarz_resolvent}{A.~Schwarz, 
{\em The partition function of a degenerate quadratic functional
and Ray-Singer invariants}, Lett. Math. Phys, {\bf 2} (1978), 247--252.} 

\bibitem{Schwarz_resolvent2}{A.~Schwarz, 
{\em The partition function of a degenerate functional}, 
Commun. Math. Phys, {\bf 67} (1979), 1--16.} 

\bibitem{Adams_Sen}{D.~H. Adams, S.~Sen, 
{\em Partition Function of a Quadratic Functional and Semiclassical 
Approximation for Witten's 3-Manifold Invariant}, hep-th/9503095.}

\bibitem{Blau_Thompson}{M.~Blau, G.~Thompson, 
Ann. Phys {\bf 205}(1991) 130.}

\bibitem{GK}{ J.~Gegenberg, G.~Kunstatter, 
{\em The Partition Function for Topological Field Theories}, 
Ann. Phys. {\bf 231} (1994) 270-289, hep-th/9304016.}

\bibitem{bv}{C.~I.~Lazaroiu, R.~Roiban and D.~Vaman, 
{\em Graded Chern-Simons field theory and graded topological
D-branes}, hep-th/0107063.}

\bibitem{Adams_Sen0}{D.~H. Adams, S.~Sen, 
{\em Phase and scaling properties of determinants
arising in topological field theories},
Phys. Lett. {\bf B 353} (1995) no. 4, 495--500, hep-th/9506079.}

\bibitem{Bismut_Lott}{J.~M.~Bismut and J.~Lott, {\em Flat vector
      bundles, direct images and higher analytic torsion}, J. Amer.
    Math Soc {\bf 8} (1992) 291.}  

\bibitem{Quillen}{D.~Quillen, {\em
    Superconnections and the Chern character}, Topology, {\bf 24},
    No.1.(1085), 89-95.} 

\bibitem{Vafa_cs}{C.~Vafa, {\em Brane/anti-Brane Systems and $U(N|M)$
      Supergroup}, hep-th/0101218.}

\bibitem{supergroupCS}{ J.~H.~Horne, {\em Skein
    relations and Wilson loops in Chern-Simons gauge theory},
    Nucl. Phys. {\bf B334} (1990) 669; Bourdeau, E.J.  Mlawer,
    H. Riggs and H.J. Schnitzer, {\em The quasirational fusion
    structure of $SU(M|N)$ Chern-Simons and W-Z-W theories}, Nucl.
    Phys. {\bf B372} (1992) 303; L. Rozansky and H. Saleur, {\em
    Reidemeister torsion, the Alexander polynomial and $U(1,1)$
    Chern-Simons theory}, J. Geom. Phys. {\bf 13} (1994) 105.}

\bibitem{superpot}{C. I.. Lazaroiu, {\em String field theory and brane 
superpotentials}, hep-th/0107162.}

\bibitem{DeWitt}{B.~DeWitt, {\em
    Supermanifolds}, Cambridge Univ. Press, Cambridge, 1984.}

\bibitem{Berezin}{F.~A.~Berezin, {\em The Method of Second
      Quantization}, Pure and Appl. Phys. {\bf 24}, Academic Press,
    New -York, 1966.} 

\bibitem{Rogers}{A.~Rogers, {\em A global theory of supermanifolds},
    J.~Math.~Phys {\bf 21} (1980) 1352-1365.}

\bibitem{Ray}{D.~B.~Ray, {\em Reidemeister torsion and the Laplacian 
on lens spaces}, Adv. in Math. {\bf 4} (1970), 109--126.}

\bibitem{RS1}{D.~B.~Ray, I.~M.~Singer, 
{\em R-torsion and the Laplacian on Riemannian manifolds}, 
Adv. Math {\bf 7} (1971), 145-210.}

\bibitem{RS_symp}{D.~B.~Ray, I.~M.~Singer, {\em Analytic torsion}, 
Proceedings of Symposia in Pure Mathematics, vol. {\bf 23}, p 167, 
American Mathematical Society 1973.}


\end{thebibliography}
\end{document}